\documentclass[%
 reprint,
 amsmath,amssymb, 
 aps,
]{revtex4-2}

\usepackage{graphicx}% Include figure files
\usepackage{dcolumn}% Align table columns on decimal point
\usepackage{bm}% bold math
\usepackage{color}
\begin{document}
\newcommand{\ojo}[1]{\textcolor{black}{#1}}
\newcommand{\ojoo}[1]{\textcolor{black}{#1}}

\preprint{APS/123-QED}

\title{Fine-Tuning of Colloidal Polymer Crystals by Molecular Simulation}% Force line breaks with \\

\author{Miguel Herranz, Clara Pedrosa, Daniel Martínez-Fernández, Katerina Foteinopoulou, Nikos Ch. Karayiannis$^*$ and Manuel Laso}
%\author{ }%
 \email{n.karayiannis@upm.es} \email{manuel.laso@upm.es}

\affiliation{%
 Institute for Optoelectronic Systems and Microtechnology (ISOM) and Escuela Técnica Superior de Ingenieros Industriales (ETSII), Universidad Politécnica de Madrid (UPM) C/ Jose Gutierrez Abascal 2, 28006 Madrid (Spain)
}%

\date{\today}% It is always \today, today,

\begin{abstract}
\par Through extensive molecular simulations we determine a phase diagram of attractive, flexible polymer chains in two and three dimensions. A surprisingly rich collection of distinct crystal morphologies appear, which can be finely tuned through the range of attraction. In three dimensions these include the face centered cubic, hexagonal close packed, simple hexagonal and body centered cubic crystals and the Frank-Kasper phase. A simple geometric model is proposed, based on the concept of cumulative neighbours of ideal crystals, which can accurately predict most of the observed structures and the corresponding transitions. The attraction range can thus be considered as an adjustable parameter for the design of colloidal polymer crystals with tailored morphologies. 
\end{abstract}
%Use showkeys class option if keyword
\maketitle

%\tableofcontents

\section{\label{sec:level1}Introduction\protect\\}

\par Crystallization is one of the most intriguing physicochemical processes in science and technology. While of paramount importance in materials design and engineering, key aspects of the phenomenon remain rather poorly understood. Thus, it is imperative to establish a connection between behavior at the level of atoms and molecules, the ensuing ordered structures and eventually the macroscopic properties of the end material. 
\par Very recently through the emergence of “digital alchemy” \cite{RN1814} the concept is tackled through a new perspective: one starts from the target morphologies and searches, mainly through computational tools, for the molecular shape and size that would produce them \cite{RN1895}. Such reverse engineering methods, combined with robust algorithms, machine learning and predictive modeling have led to significant advances in the computer-aided design of soft materials made of self-assembled colloids and nanoparticles \cite{RN116, RN1896}. As an alternative, the shape/size of hard-body objects can be effectively replaced by fine tuning the pair-wise interactions between the species, be particles or atoms,  in order to achieve the desired geometric patterns \cite{RN1898, RN1897, RN1899}. These simulation breakthroughs have been accompanied by vigorous progress in colloidal synthesis allowing for a systematic, instead of trial-and-error, fabrication of optimal structures based on self organization from properly selected building blocks \cite{RN1819, RN1871, RN1872, RN439, RN1818, RN1817}.
\par The phase behavior of macromolecular systems is equally important \cite{RN1462} and more complex compared to the one of monomeric analogs due to the wide spectrum of characteristic length and time scales involved. Designing crystals made of soft/hard colloidal polymers and molecules \cite{RN1875, RN1873} remains a formidable challenge in spite of the important experimental \cite{RN1877, RN1879, RN1880, RN1874, RN1878, RN1876, RN377} and modeling \cite{RN1558, RN1748, RN1726, RN1339, RN1809} advances in synthesis, characterization and property-prediction. 
\par In the present contribution we demonstrate a hierarchical modeling approach for the design and morphological fine tuning of crystals of colloidal polymers with short-range attractive interactions. First, we describe the macromolecular system at hand, then we introduce a geometric neighbor model to predict the thermodynamically stable crystal,  and finally we resort to simulations to verify and extend analytical predictions. 
\section{\label{sec:level2}Model, Systems and Methodology\protect\\}

\subsection{Molecular model}

\par \ojoo{The colloidal polymer model we have chosen (in two and three dimensions) is a bulk assembly of linear, freely-jointed chains of tangent, non-overlapping spheres of uniform diameter $\sigma_1$, which is further the characteristic unit length of the system, taken here as unity. Bonded monomers along the chain backbone are tangent within a numerical tolerance of $dl=6.5\times 10^{-4}$. Practically, this means that no gaps exist between bonded sites that are known to profoundly affect the phase behavior of chains and the ensuing crystal morphologies \cite{RN497, RN319}. Short-range, pair-wise attraction is realized through the square well (SW) potential, described by} 

\ojoo{
\begin{equation}
u_{SW}(r_{ij}) =
\begin{cases}
0, r_{ij} \ge    \sigma_2 \\
-\epsilon, \sigma_1 \le r_{ij} < \sigma_2 \\
\infty,    r_{ij} < \sigma_1\
\end{cases}      
\label{USW}     
\end{equation}
}

\ojoo {where the tunable parameters correspond to intensity (depth) $\epsilon$ and the range  of interaction $\sigma_2$, the latter being expressed in units of $\sigma_1$ ($\sigma_2>\sigma_1$). In Eq. \ref{USW}, $v_{SW}(r_{ij})$ is the energy that corresponds to the interaction of two monomers $i$ and $j$ whose centers lie at a distance $r_{ij}$.
}
\par Since the early works of Young and Alder \cite{RN1528, RN1529} the SW model has been applied to study (free) energy-driven processes in monomeric and more complex systems, and has provided insights into the phase behavior, coalescence and percolation of monomers \cite{RN1540, RN1539, RN1271, RN507, RN1273, RN1537, RN1536, RN1285, RN1283}, and protein folding and self-assembly of single chains \cite{RN1329, RN1327, RN1328, RN1440, RN1284}. 
\subsection{Geometric Neighbor Model}

\par The proposed geometric  neighbor model is particularly simple and it follows a concept similar to the one proposed by Serrano-Illán $et$ $al$. as analyzed in \cite{RN1536}. Given the entirely attractive nature of the interactions, thermodynamic stability of the crystals is primarily dictated by the number of neighbours as a function of distance. The more neighbours packed within a radius equal to the attraction range $\sigma_2$, the lower the potential energy and (ignoring entropic differences among polymorphs, which are known to be quite small \cite{RN125, RN2011, RN132, RN441, RN1735}) the more stable the corresponding crystal. 
\par The tangency condition imposed by chain connectivity simplifies the analysis of polymeric analogs. First, because only strictly tangent polymer crystals have to be considered. Second, the bonds along the chain backbone stabilize cluster formation compared to monomeric analogs for the same attraction intensity.
\par The following crystals have been considered for the geometric neighbor model: honeycomb (HON), square (SQU) and triangular (TRI) 2-D crystals,  and the face centered cubic (FCC), hexagonal close packed (HCP), holoedric 6/mmm (simple hexagonal or HEX), and body centered cubic (BCC) 3D crystals, whose structures and properties can be found in \cite{RN1542}.
\par
\begin{figure}
\centering

\includegraphics[scale=1.0]{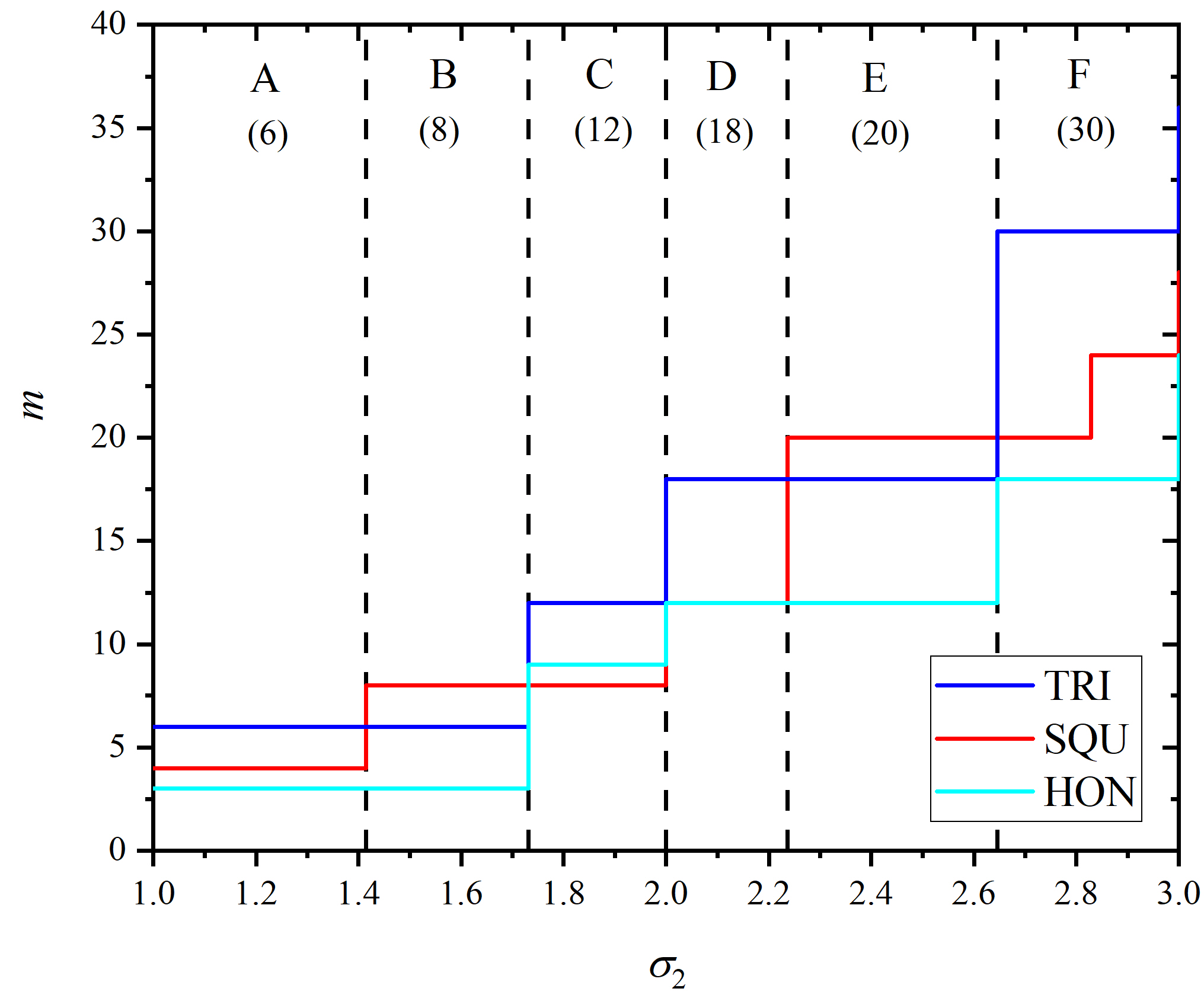}
\includegraphics[scale=1.0]{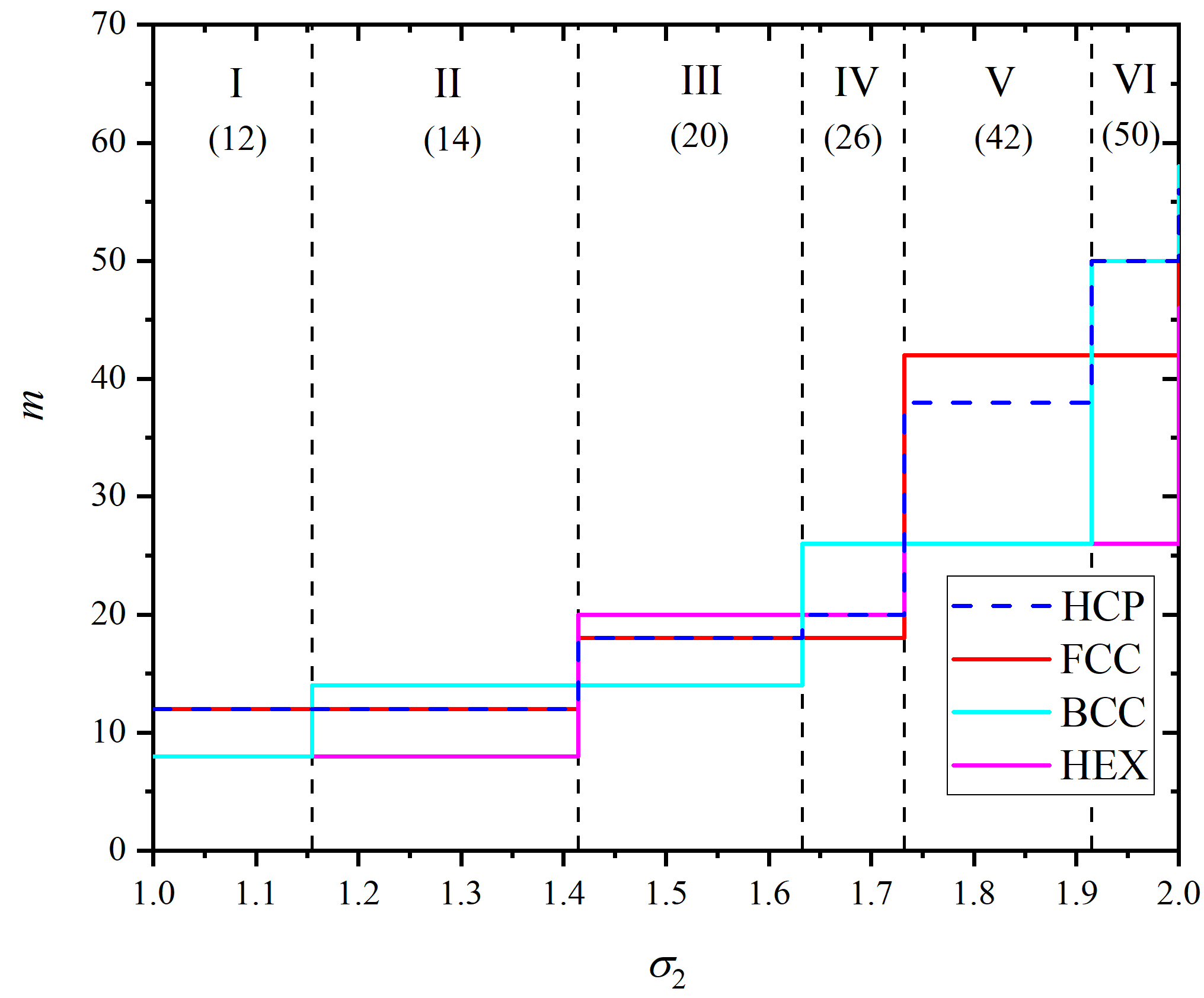}

\caption{Number of neighbours, $m$, as a function of distance from a reference site, $\sigma_2$, for (bottom panel) the HCP, FCC, BCC and HEX crystals in 3D and (top panel) the TRI, SQU and HON crystals in 2D. Dashed vertical lines identify the values at which the steps take place. Each region is identified by a roman number (3D) or letter (2D).  Arabic numbers in parentheses correspond to the cumulative number of neighbours for the most stable crystal in the given region.}
\label{models}
\end{figure}

 Fig.\ref{models} shows the number of neighbours as a function of distance for the above crystal types in 2D (top panel) and 3D (bottom panel).  Starting with 2D, the geometric model predicts that the TRI crystal is the prevailing one from the point of view of energy in all ranges except the intervals $1.41 \leq r \leq 1.73$ (Region B) and $2.24 \leq r \leq 2.65$ (Region E) where a site in the SQU crystal has more neighbours within a circle of diameter $\sigma_2$. The HON is systematically more dilute and thus the least stable crystal. 
\par In the 3D case, six distinct regions can be identified. The initial dominance of HCP/FCC (Region I) is succeeded by BCC (Region II), HEX (Region III), again BCC (Region IV), FCC (Region V) and HCP/BCC (Region VI). At a first glimpse it is quite surprising that for distances in the range of $1.15 \leq r \leq 1.73$ the non-compact HEX and BCC crystals prevail. Still, the dominance of non-compact crystals, as calculated here, is in agreement with the density-based calculations in Ref \cite{RN1536}.
\par The phase diagram of Fig. \ref{models}, as predicted by the geometric neighbor model,  raises a number of intriguing questions, especially whether the expected ordered morphologies can appear “in reality”. We remind here that any entropic contributions, both translational and conformational, are ignored. Information in the geometric neighbor model about chain connectivity is solely incorporated through the tangency condition, effectively enforcing the absence of gaps between the inter-lattice sites.  Thus, this straightforward geometric argument can only be considered as a first order approximation (but an exceedingly successful one, as will be shown below).
\subsection{Monte Carlo Simulations}

\par To validate the predictive capacity of the proposed model, we carry out extensive simulations for the generation and equilibration of the systems composed of 100 attractive chains of average chain length $N_{av} = 12$ for a total of $N = 1200$ interacting sites. \ojoo{Regarding the simulations and the successive structural identification of the generated system configurations, we employ Simu-D, a home-made simulator/descriptor software suite \cite{RN1824}. The simulator component is a Monte Carlo (MC) protocol based on local, cluster and chain-connectivity-altering moves (CCAMs) \cite{RN9,RN16,RN1252} as in all our recent works on (free)energy-driven \cite{RN1541} and athermal \cite{RN1678,RN1894,RN2010} polymer-based systems.
The MC mix is composed of: (i) rotation (10\%), (ii) reptation (10\%), (iii) flip (34.7\%), (iv) intermolecular reptation (25\%), (v) configurational bias (20\%), (vi) simplified end-bridging (0.1\%), (vii) simplified intermolecular end-bridging (0.1\%) and cluster moves (0.1\%) where numbers in parenthesis correspond to attempt probabilities. Furthermore, a cluster analysis is attempted at regular intervals ($10^7$ steps) to disable cluster moves when there is just one cluster. Clusters are detected using an approach similar to the DBSCAN algorithm \cite{confkddEsterKSX96, RN2012}, where the distance criterion is set to 1.2 (in units of $\sigma_1$). As explained in detail in \cite{RN16, RN1252, RN1824} every local MC move is executed in a configurational bias pattern. To increase computational performance the number of attempts depends on cluster population, being $n_{dis} = 5$ when there is just one cluster and $n_{dis} = 20$ when several exist. }

\par \ojoo{Simulations are carried out in NVT ensemble for 3D simulations, and in the NPT ensemble for 2D. In both cases, temperature is set equal to $T = 1 / k$, where $k$ is the Boltzmann constant. Pressure is fixed at 1 bar (NPT simulations) and packing density at $\varphi = 0.05$ (or equivalently number density  $ \rho_{n} = 0.0262$) (NVT simulations).  Due to the application of chain-connectivity-altering moves, dispersity in chain lengths is introduced. These vary uniformly with the minimum and maximum allowed lengths being set at 6 and 18 monomers, respectively. Selected simulations on polymers with longer lengths ($N$ = 24), under the same conditions, revealed no appreciable difference in the phase behavior. }

\par \ojo{We start from a fully equilibrated athermal system of fully flexible chains of tangent hard spheres at low packing density and activate the SW potential. As explained in detail in \cite{RN1541}, this corresponds practically to instantaneous quenching with the attraction intensity adopting the role of an effective quench rate: the higher the value of $\epsilon$, the higher the temperature difference.} \ojoo{2D polymer configurations are created by shrinking the original 3D systems until a film thickness of unit length is reached through a process that is described in \cite{ClaraP}. By fixing all other parameters ($T$, $P$ or $V$ and $dl$) and the interaction intensity at $\epsilon = 1.2$, we systematically explore the effect of interaction range in the interval $\sigma_2 \in [1.10,...., 2.00]$ in steps of 0.01 for 3D systems and in steps of 0.02 for 2D systems. As a first design step, a representative value of the well depth $\epsilon = 1.2$ was selected because it is a) large enough to prevent entropy from overwhelming the internal energy advantage of specific crystals, and b) not so high that the simulations would become trapped in local energy minima, leading to glass formation instead of crystallization (see for example Figure 5 in \cite{RN1541} on the phase behavior as a function of attraction intensity).}
\par \ojo{Equilibration of the systems is traced through a hierarchical, two-step evolution: 1) A single cluster is formed including all chains and their monomers. Quantification of this step is trivial by tracking the evolution of the number of formed clusters as the simulation evolves. Activation of cluster-based MC moves is a necessity especially for low values of $\sigma_2$ and/or high values of $\epsilon$ \cite{RN1541}; 2) The state of order in the formed cluster becomes stable. This final step is quantified through the evolution of the degree of crystallinity, practically being equal to the sum of all order parameters for all reference crystals (see next section). To check the reproducibility of the presented results we have conducted additional simulations starting from different initial athermal configurations (but still under very dilute conditions) and with different seeds for the random number generators. For the whole range of studied $\sigma_2$ values no appreciable difference is detected in the established morphologies between independent MC simulations.}

\subsection{Structural Analysis of Computer-Generated System Configurations}

\par \ojoo{We employ the Characteristic Crystallographic Element (CCE) norm \cite{RN10, RN1542} for the structural analysis of the computer-generated system configurations. In 3D, we use the CCE norm to detect hexagonal close packed (HCP), face centered cubic (FCC), simple hexagonal (HEX), and body centered cubic (BCC) crystals as well as non-crystallographic fivefold (FIV) local symmetry. Regarding the 2D analysis, we compare against the triangular (TRI), square (SQU), and honeycomb (HON) crystals, as well as pentagonal (PEN) local symmetry. Every monomer in the system, be in 2D or 3D, is tested against all corresponding reference crystals as described above. Accordingly, a norm value $\mu_{j}^{X}$ is obtained for site $j$ with respect to a reference crystal $X$. Here, a threshold value of $\mu^{c}=0.245$ is adopted below which a site is considered as of $X$-type. Furthermore, we can measure the order parameter, $S^X$ of a given $X$ crystal as:}

\ojoo{
\begin{equation}
S^X = \int_{0}^{\mu^{c}} P(\mu^{X}) \,d\mu^{X} \ 
\label{CCE_order_par}     
\end{equation}
}

\ojoo{where $P(\mu^{X})$ is the probability distribution function over all monomers for a given configuration. In order to avoid considering the monomers that lie on the surface of the formed cluster the value of $S^X$ is multiplied by  $\frac{N}{N-N_{surf}}$, where $N$ is the number of sites in the simulation ($N = 1200$ in all simulations reported here) and $N_{surf}$ is the is the number of monomers on the outer surface of the formed cluster. This is because in 3D under dilute conditions surface monomers lack a complete Voronoi environment and thus are characterized by a highly disordered local structure.}

\section{\label{sec:level3}Results\protect\\}
\par The activation of attractive potential dictating all intra- and inter-chain interactions leads to aggregation of the polymer chains and to the eventual formation of a single cluster, which may crystallize depending on the values of interaction range and intensity. As stated earlier the selected value of $\epsilon = 1.2$ under the specific simulation conditions guarantees crystallization over glass formation as demonstrated in \cite{RN1541}. Once the single cluster containing all chains and monomers is formed, the CCE descriptor is employed to quantify the structural characteristics and the possible similarity to one of the reference crystals in 2D or 3D. \ojoo{Throughout the manuscript the following color convention is used: Blue, red, green, pink, and cyan colors correspond to HCP, FCC, FIV, HEX, and BCC similarity, respectively, for 3D systems; Blue, red, green, and cyan colors correspond to TRI, SQU, PEN, and HON similarity, respectively, for 2D systems. Amorphous (or more precisely unidentified or "none of the above") sites (AMO) are shown in yellow (and with reduced dimensions in 3D for clarity).}   

\par
\begin{figure}
\centering

\includegraphics[scale=0.5]{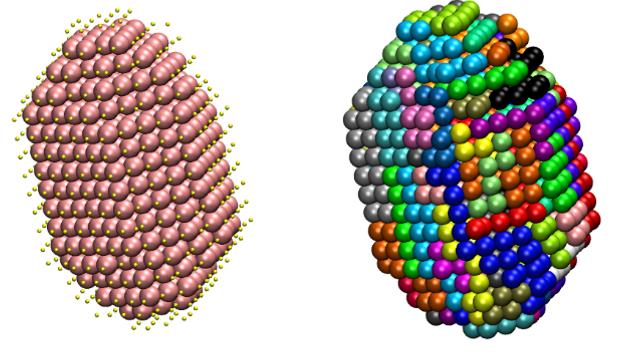}

\caption{Final stable polymer cluster as obtained from NVT simulations in 3D with interaction range set at $\sigma_2 = 1.51$. Left panel: Spheres are color-coded according to their structural similarity: Blue, red, green, pink and cyan colors correspond to HCP, FCC, FIV, HEX, and BCC similarity, respectively. Amorphous (AMO) sites are colored in yellow with reduced dimensions for clarity. Right panel: Spheres are colored according to their parent chain. Snapshots created with the VMD software \cite{RN250}.}
\label{cluster_3D}
\end{figure}

\begin{figure}
\centering

\includegraphics[scale=0.4]{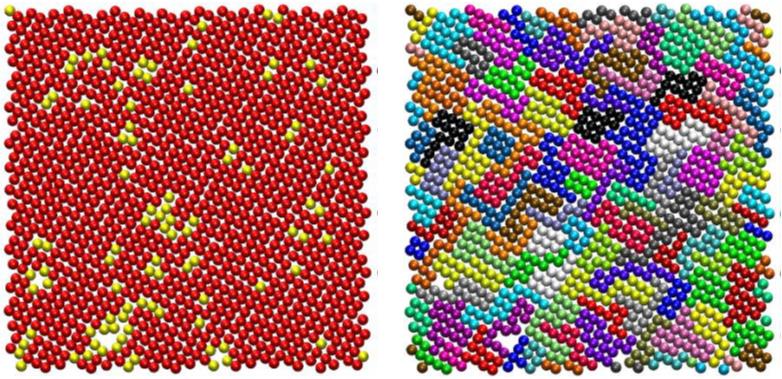}

\caption{Final stable polymer cluster as obtained from NPT simulations in 2D (polymer films of thickness equal to $\sigma_1 = 1$ with interaction range set at $\sigma_2 = 1.58$). Left panel: Spheres are color-coded according to their structural similarity: Blue, red, green, and cyan colors correspond to TRI, SQU, PEN, and HON similarity, respectively. Amorphous (AMO) sites are colored in yellow. Right panel: Spheres are colored according to their parent chain. Snapshots created with the VMD software \cite{RN250}.}
\label{cluster_2D}
\end{figure}

 \ojoo{An example of the formed single cluster of polymers at the end of the simulation and its structural identification through the CCE norm descriptor can be seen in Figs. \ref{cluster_3D} and \ref{cluster_2D}, for 3D and 2D systems, respectively. On the left panel sites are colored according to their structural type, as quantified by the crystallographic analysis, and on the right panel monomers are colored according to the parent chain.}

\par In the literature, simulation results are typically reported in reduced units of temperature $(T^*=(k T)/\epsilon)$ and pressure $(P^*=(P\sigma_1^3)/\epsilon)$, where $T$ and $P$ are the applied temperature and pressure and $k$ is Boltzmann's constant. Here, we deviate from the traditional approach and examine the important dependence on the attraction intensity,  which, unlike $T$ or $P$, is a material dependent, adjustable parameter, in accordance with our goal to finely tune specific morphologies of hard colloidal polymers.  

\par
\begin{figure}
\centering

\includegraphics[scale=1]{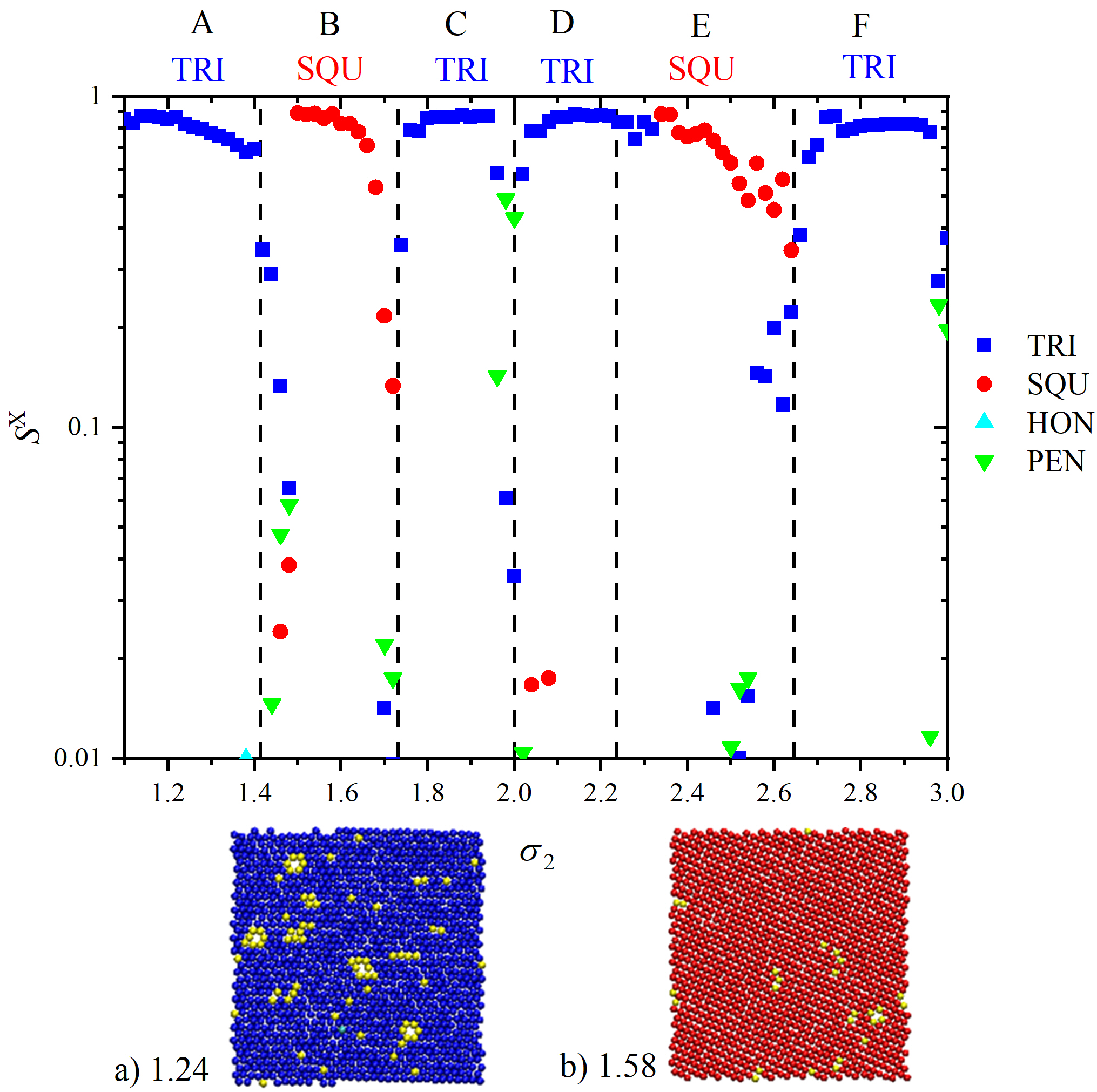}

\caption{Fraction of sites with a similarity to a given reference crystal or local symmetry $X$, $S^X$, as a function of the interaction range, $\sigma_2$, as obtained from MC simulations of attractive polymer chains in 2D (films of thickness equal to $\sigma_1$). Distinct regions, the thresholds and the corresponding prevailing crystals, as predicted by the simple geometric neighbour model, are identified by the roman letters, the dashed vertical lines and the color labels, respectively. Also shown are system configurations at the end of the simulation for selected values of the interaction range. Spheres  are color-coded according to their structural similarity: Blue, red, cyan, and green colors correspond to TRI, SQU, HON, and PEN similarity, respectively, as quantified by the CCE norm analysis \cite{RN1542}. Amorphous (AMO) sites are colored in yellow. Snapshots are created with VMD software \cite{RN250}.} 
\label{2dphasediagram}
\end{figure}

 Fig. \ref{2dphasediagram} illustrates the strong effect, and hence the great tunability potential, of the interaction range~$\sigma_2$: the fraction of crystalline sites is shown as a function of $\sigma_2$ for the 2D systems, together with the regions predicted by the simple geometric neighbour model and the expected prevailing crystals, based on energetic considerations only.  Model predictions and simulation results are in excellent agreement, especially if we consider the simplicity of the model reported above.  Not only is the agreement between the predicted and observed crystal types striking, but also the precision in the values of the interaction range $\sigma_2$ at which polymorph transitions take place over the whole interaction range.
\par The immediate conclusion that can be drawn is that, for the reasonable well depth $\epsilon = 1.2$, entropy differences between polymorphs play a very subordinate role in the selection of a particular polymorph type.  This is in agreement with the known small differences in entropy among crystals of monomeric spheres \cite{RN125, RN2011, RN132, RN441, RN1735}, and also with recent quantitative estimates of differences in chain conformational entropy among crystals of polymers of hard spheres \cite{RN1894,MiguelH}.

\begin{figure}
\centering
\includegraphics[scale=0.28]{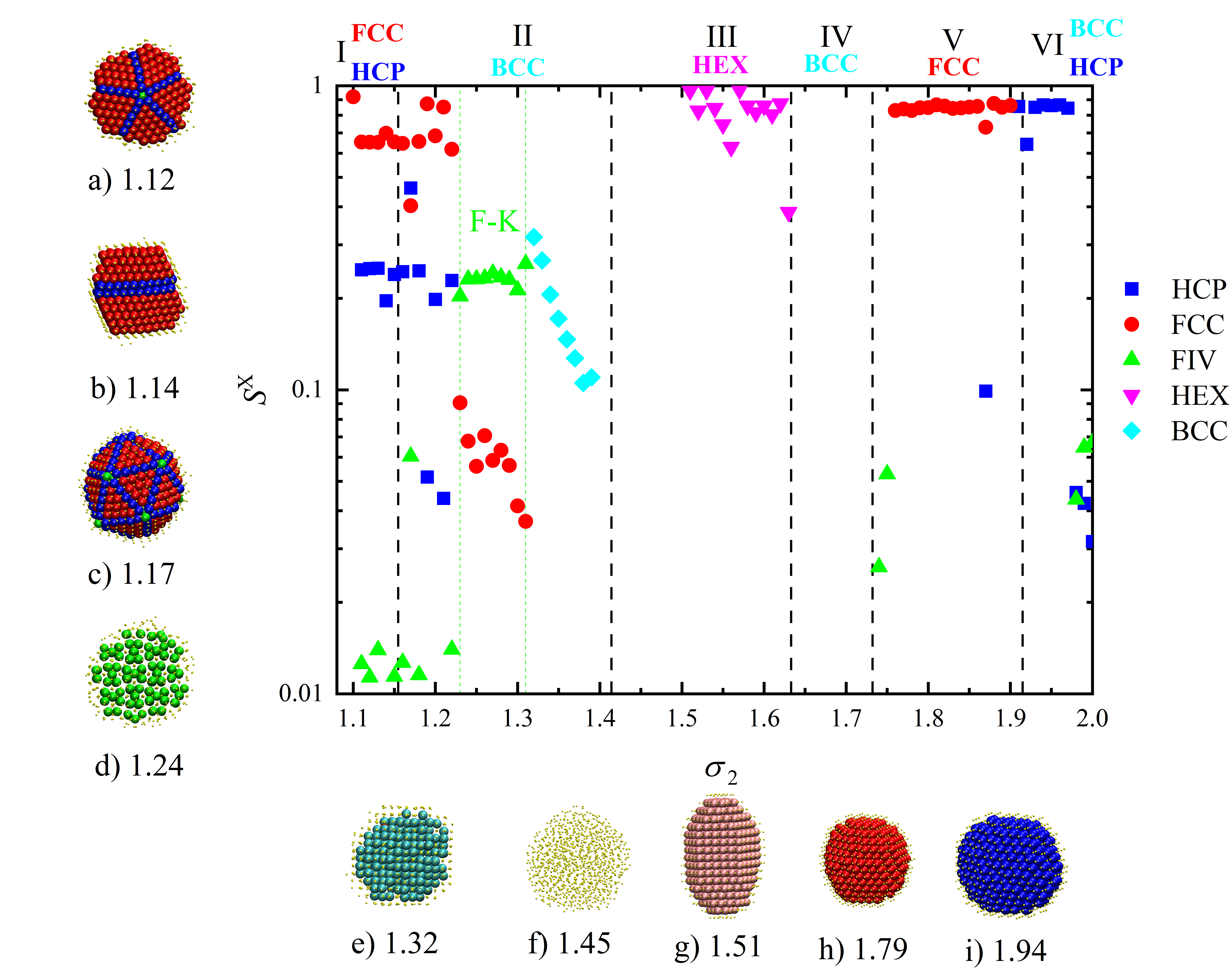}
\caption{Fraction of sites with a similarity to a given reference crystal as a function of the interaction range, as obtained from MC simulations on attractive polymer chains in bulk 3D. Outer surface monomers are excluded in the calculation of $S^X$. Distinct regions, the thresholds and the corresponding prevailing crystals, as predicted by the geometric neighbour model, are identified by the roman numbers, the dashed vertical lines and the color labels, respectively. Also shown are system configurations at the end of the simulation for selected values of the interaction range. Spheres are color-coded according to their structural similarity: Blue, red, green, pink, and cyan colors correspond to HCP, FCC, FIV, HEX, and BCC similarity, respectively. Amorphous (AMO) sites are colored in yellow with reduced dimensions for clarity. The regime of the Frank-Kasper (F-K) phase, as established in simulations, is indicated by the vertical green dashed lines. Snapshots are created with VMD software \cite{RN250} and are also available in \ref{snapshots}, and for better visual inspection as 3-D, interactive images in Supplementary Material (S.M.) \cite{Suppl}.}
\label{3dphasediagram}
\end{figure}
\par
\begin{figure*}
\centering
\includegraphics[scale=0.8]{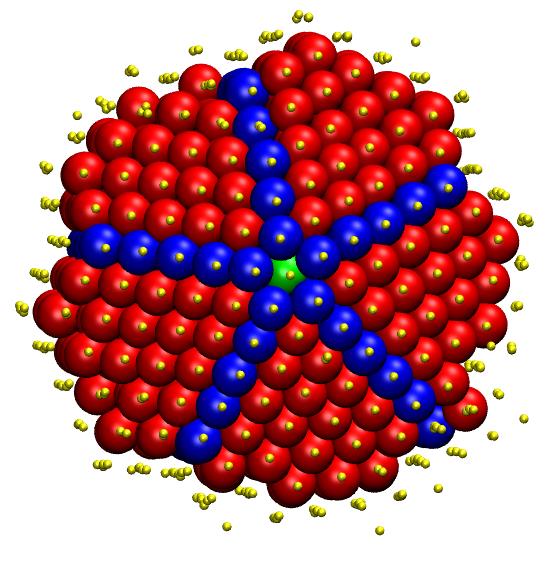}
\includegraphics[scale=0.8]{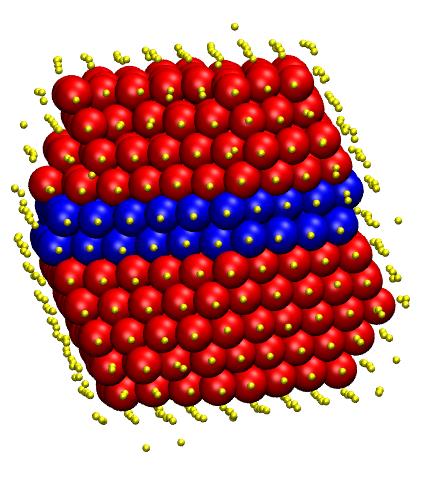}
\includegraphics[scale=0.8]{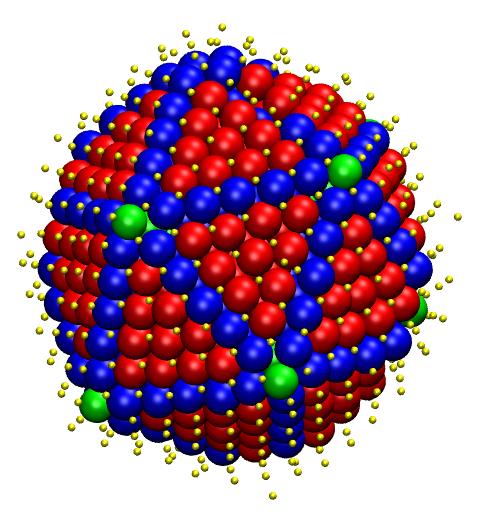}
\includegraphics[scale=0.8]{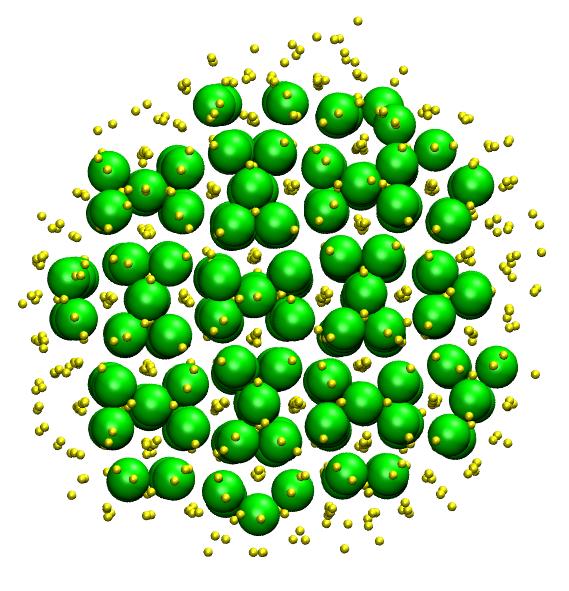}
\includegraphics[scale=0.8]{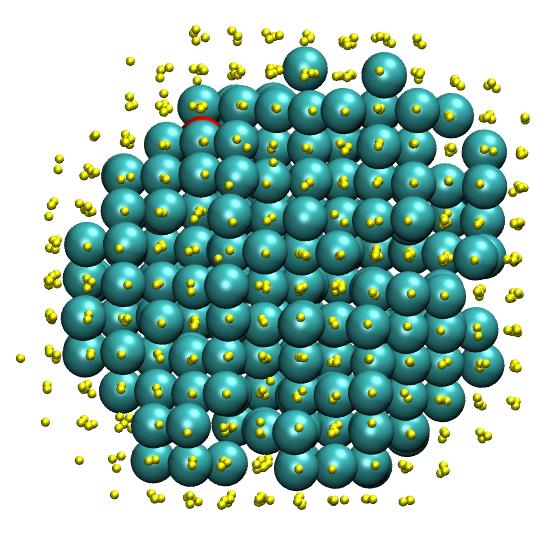}
\includegraphics[scale=0.8]{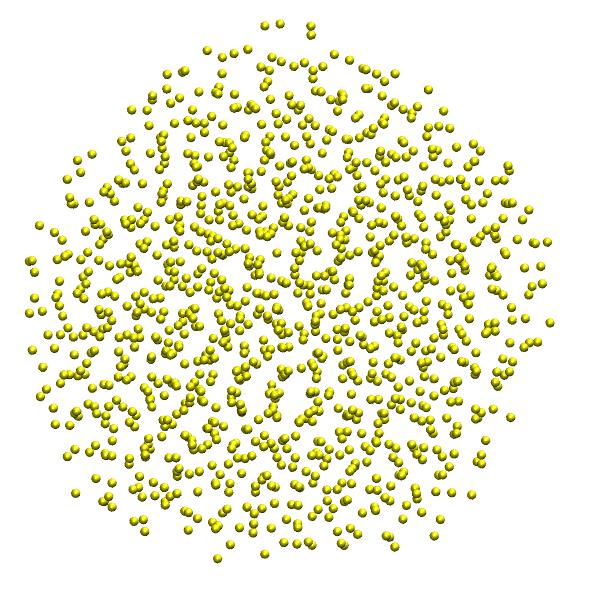}
\includegraphics[scale=0.8]{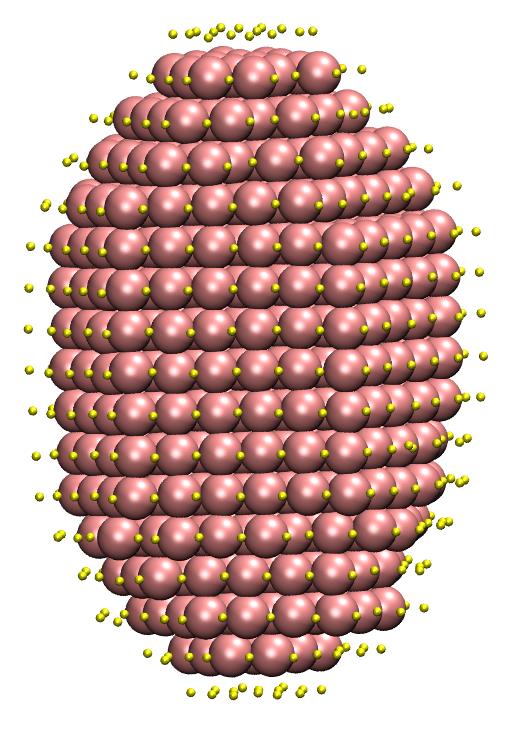}
\includegraphics[scale=0.8]{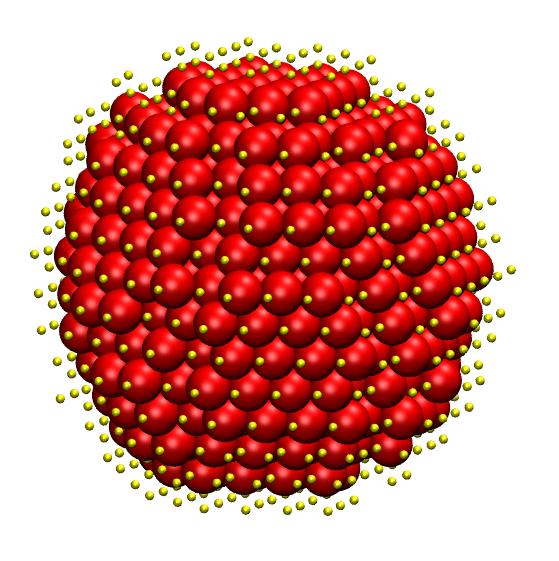}
\includegraphics[scale=0.8]{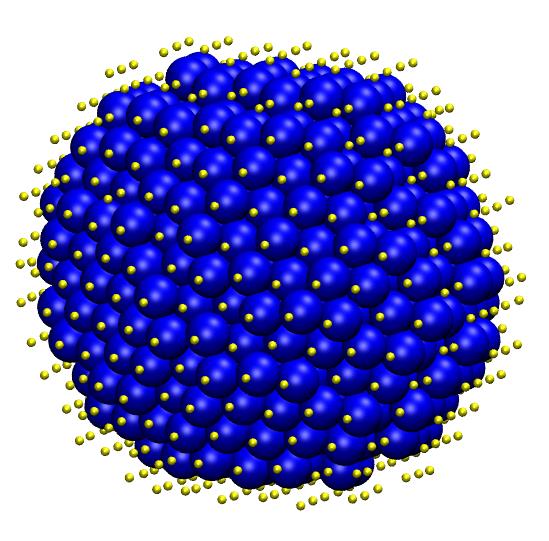}
\caption{Representative system configurations (bulk 3D) at the end of the Monte Carlo simulations with varied attraction range, $\sigma_2$. All chains have formed a single cluster of varied order and morphology. Top: $\sigma_2 = 1.12$ (random hexagonal close packed, rHCP), 1.14 (rHCP), 1.17 (rHCP), 1.24 (Frank-Kasper, F-K) and 1.32 (BCC); Bottom: $\sigma_2 = 1.45$ (amorphous, AMO), 1.51 (HEX), 1.79 (FCC), and 1.94 (HCP). Labels in parenthesis correspond to the established structures/morphologies. Spheres are color-coded according to their structural similarity: Blue, red, green, pink, and cyan colors correspond to HCP, FCC, FIV, HEX, and BCC similarity, respectively. Amorphous (AMO) sites are colored in yellow with reduced dimensions for clarity. All clusters have the same number of monomers ($N = 1200$), their size may appear different because of the zoom level and the different viewing angle. Snapshots are created with VMD software \cite{RN250} and are also available for visual inspection as 3-D, interactive images in Supplementary Material (S.M.) \cite{Suppl}.}
\label{snapshots}
\end{figure*}

 Fig. \ref{3dphasediagram} presents a similar phase diagram for the 3D systems, along with the predictions of the neighbour model, while Fig. \ref{snapshots} hosts representative snapshots at the end of the MC simulations. At short interaction range (Region I) a random hexagonal close packed (rHCP) crystal is observed. The rHCP morphology can have a unique stacking direction of fivefold-free HCP and FCC layers, or multiple stacking directions where the meeting (composition) planes at the crystal boundaries are fivefold-ridden. As the number of neighbours is the same between HCP and FCC no pure crystal prevails and, thus, the rHCP polymorph remains the final ordered morphology in Region~I. The rHCP dominance, as gauged by simulations, extends into higher values of $\sigma_2$ than predicted by the neighbor model. 

\par
\begin{figure}
\centering
\includegraphics[scale=0.50]{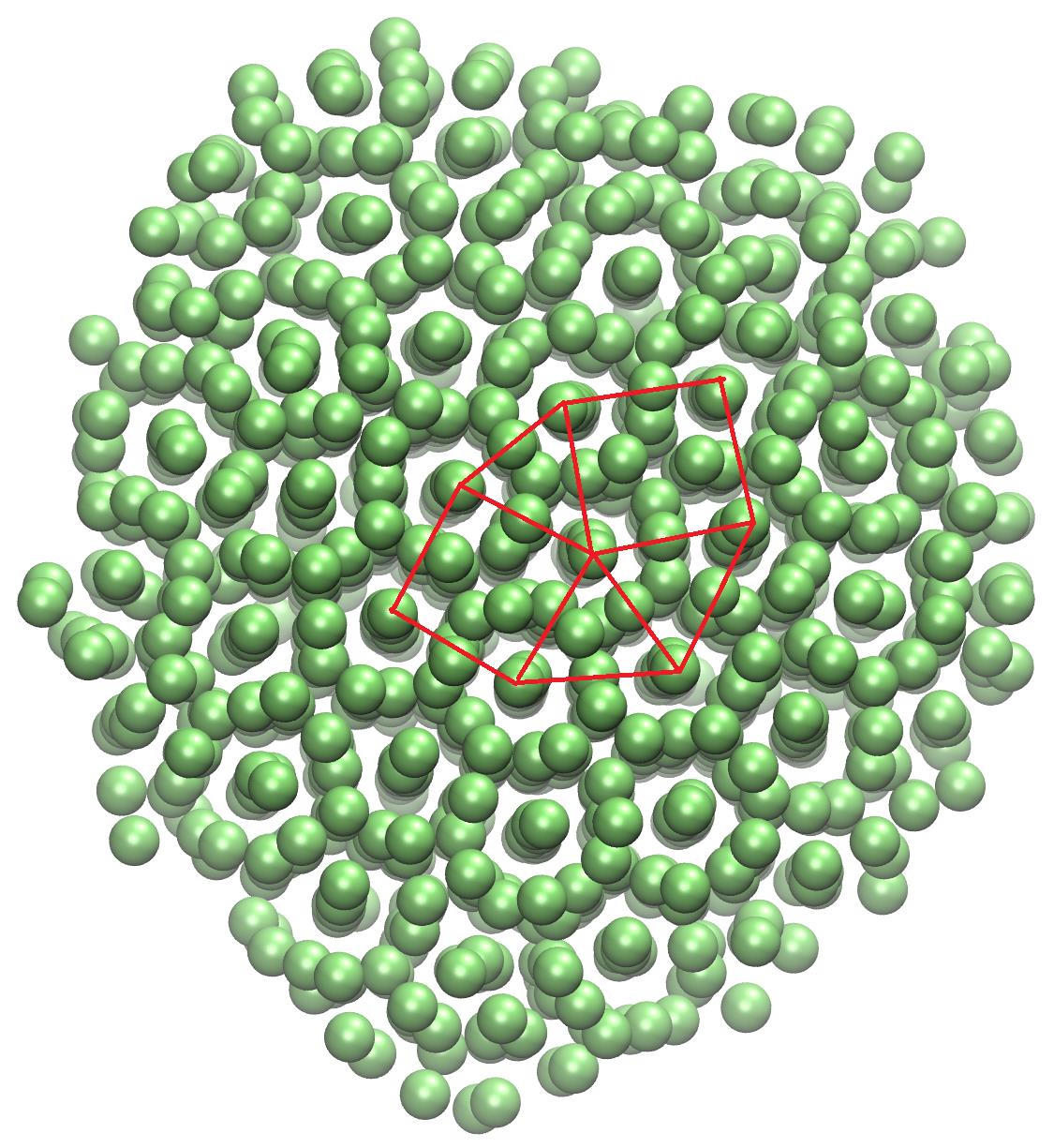}
\caption{System snapshot at the end of the MC simulation at $\sigma_{2} = 1.23$ corresponding to the formation of the $\sigma$ variant of the Frank-Kasper phase. Red lines are indicative of the tiling pattern of squares (`4') and triangles (`3') applied on the sparsely populated layer of spheres corresponding to the $3^2.4.3.4$ format. The lime color used here for the spheres is not to be confused with the green color to indicate FIV similarity through the CCE-norm description.} 
\label{Frank-Kasper}
\end{figure}

 Interestingly, in the region between $1.21 \leq \sigma_2 \leq 1.30$ none of the crystals, expected by the geometric model, appears. The resulting structure is characterized by an abundance of fivefold sites and the absence of any appreciable population of sites with crystal similarity as captured by the CCE-norm descriptor. Close inspection of the established morphology, as hosted in Fig. \ref{Frank-Kasper}, reveals that it corresponds to the $\sigma$ variant of the Frank-Kasper (F-K) phase \cite{RN1951,RN1952}. In the past, F-K phases have been reported in studies of self organizing soft matter systems, including macromolecules, colloids, surfactants and liquid crystals \cite{RN1980, RN1981, RN1982, RN1983, RN1984, RN1985, RN1986, RN1987, RN1988, RN1989, RN1990}. Red lines in Fig. \ref{Frank-Kasper} connect triangles `3' and squares `4' in the tiling of the sparsely-populated layer of the F-K phase. The resulting tiling of $3^2.4.3.4$ is characteristic of the $\sigma$ F-K phase. 
\par The expected dominance of the imperfect BCC crystal sets in at higher values of $\sigma_{2}$, compared to the threshold predicted by the neighbour model. The transitions BCC (Region II) $\Leftrightarrow$ HEX (Region III) and HEX (Region III) $\Leftrightarrow$ BCC (Region IV) exist, as clearly captured by the data in Fig. \ref{3dphasediagram}. However, they are both accompanied by the presence of regions (\ojo{called here “amorphous zones”, denoted in the phase diagram as "AMO"}) of glassy, disordered behavior where no crystal traces are detected. The nature and origin of the two \ojo{AMO zones}, along the phase diagram, which is otherwise rich in distinct crystal morphologies, are open topics under study.  
\par The FCC and HCP prevalence and the value of $\sigma_2\approx1.91$ at the transition match very well those expected from the number of neighbours in Regions V and VI. In fact, in Regions V and VI, we can observe the formation of perfect FCC and HCP crystals,  in contrast to the behavior in Region I where the rHCP polymorph dominates. This trend can be explained rather trivially by the neighbour model: in Region I there is a tie in the number of neighbours (12) for the HCP and FCC crystals, and hence in internal energy. The tiny entropic difference between FCC and HCP \cite{MiguelH} is not sufficient to make FCC dominant in the MC simulation. However, in Regions V and VI the FCC and HCP crystals dominate by a difference of 2 and 10 neighbours, respectively, and are thus the energetically favoured states. The high number of neighbours in these two regions (38 and 50 respectively) also imply that monomers reside in a much deeper potential energy well, so that entropy plays a very minor role in selecting the stable polymorph, and in blurring the boundaries between phases. 
\begin{figure}
\centering
\includegraphics[scale=1.0]{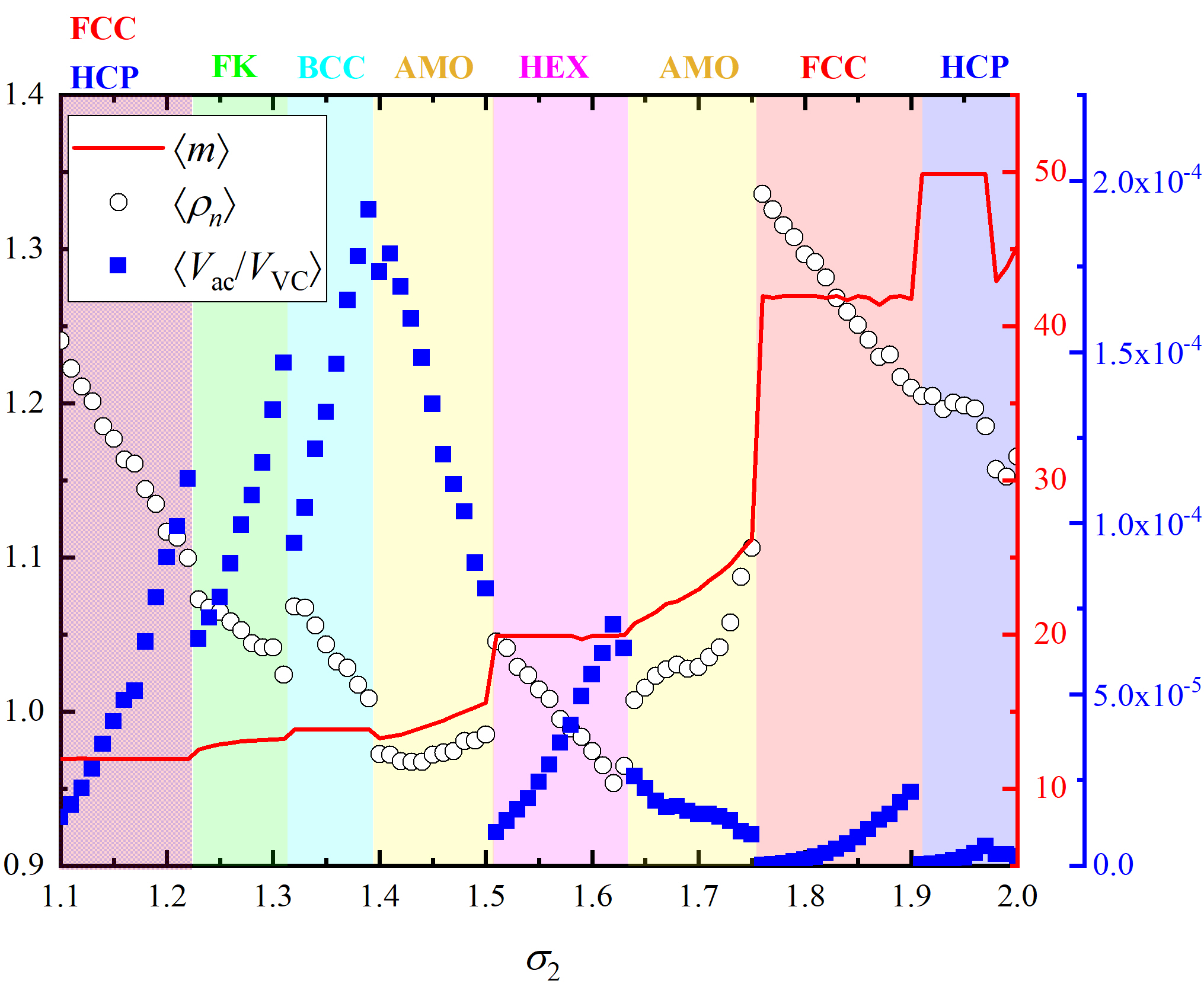}
\caption{Local number density $\langle \rho_n\rangle$ (left $y$-axis, scattered points), number of neighbours  $\langle m\rangle$ (right $y$-axis, red line) and ratio of accessible volume divided by the total volume of the Voronoi cell  $\langle V_{ac}/V_{VC}\rangle$ (right $y$-axis, blue line) inside a shell of radius $\sigma_2$ as a function of $\sigma_2$. $\langle\quad \rangle$ corresponds to average over all sites with fully developed Voronoi environment (i.e. surface monomers of the cluster are excluded from the analysis) and all snapshots that correspond to the equilibrated part of the trajectory. Background colors correspond to the dominant (crystal, F-K or amorphous) phase as observed in the MC simulations.} 
\label{densityplot}
\end{figure}
\par In the present study we let the crystal be self-assembled in almost vacuum conditions, so it can spontaneously adjust its density and structure by obeying the tangency condition, the orientational and radial symmetry and by minimizing its free energy, chiefly by maximizing the number of neighbours within a shell of size $\sigma_2$. The complex phase diagram is primarily a consequence of this maximization. In order to interpret these observations,  we plot in Fig. \ref{densityplot} the local number density, $\langle \rho_n\rangle$, the number of neighbours, $\langle m\rangle$, and the ratio of the accessible volume by the volume of the Voronoi cell, $\langle V_{ac}/V_{VC}\rangle$, as a function of interaction range. The brackets $\langle\quad \rangle$ denote here averaging over all monomers which have fully developed Voronoi environment, i.e. that do not lie on the  surface of the cluster, and over all system configurations in the final stable part of the simulation trajectory. Local number density is calculated as the inverse of the Voronoi cell volume. The Voronoi tessellation, also a requirement for the crystallographic analysis, is done through the voro++  software \cite{RN1543}. The regions observed in the MC simulations are identified in Fig. \ref{densityplot} by the different colors and the corresponding labels. Well defined crystals are characterized by constant number of neighbours in particular ranges of $\sigma_2$, in excellent qualitative agreement with the model predictions in Fig. \ref{3dphasediagram}, and by negative slope in the density curve: within the domain of a given polymorph, the crystal expands as $\sigma_2$ increases, while keeping the number of neighbours constant.
\par In the crystalline regions where the number of neighbours remains constant, the energy is also strictly constant due to the flatness of the square well potential. Therefore, in these regions minimization of free energy is tantamount to maximization of entropy,  most of which is of translational origin \cite{RN125, RN2011, RN132, RN441, RN1735} and, ignoring for simplicity the shape of the local environment,  is proportional to the volume accessible to the monomers. In parallel, an expansion of the Voronoi cell is also taking place, driven by entropy. This expansion is reflected in a higher accessible volume ($V_{ac}$), which eventually increases the translational entropy of the monomers. The accessible volume can be estimated through MC integration, considering that $V_{ac}$ is effectively the fraction of the Voronoi cell volume ($V_{VC}$) where a spherical monomer can be placed without overlapping with the walls of the enclosing polyhedron.  So long as the number of neighbours remains constant, the crystal expands with increasing $\sigma_2$ primarily because of the increase in translational entropy of the chain monomers. Chain conformational entropy changes only very weakly with $\sigma_2$ and is thus not the driving force for crystal expansion with $\sigma_2$. Fig. \ref{densityplot} shows clearly that the accessible volume increases in the crystal regions, while in the \ojo{AMO zones}, where amorphous behavior is observed, it tends to shrink. 
\par Transitions between well defined crystals, for example the FCC $\Leftrightarrow$ HCP transition, are marked by a jump in the number of neighbours and a moderate change in the (still negative) density slope.  The behavior changes drastically in the transition between well defined crystals and the \ojo{“AMO zones"} where disorder prevails. There, the amorphous cluster contracts and the number of neighbours increases in a smooth, rather than step-wise, pattern. This behavior is identical for both  \ojo{AMO zones}, which surround the domain of HEX ordered morphologies. Negative density slopes are a rather straightforward consequence of entropic pressure: as long as the sites remain within the interaction range, increasing the cell volume leads to larger translational entropy.  The expansion is then energetically neutral and entropically favorable, so that the crystal expands. As $\sigma_2$ grows beyond specific limits, the number of neighbours and thus its stability increases through a transition to another polymorph, mostly driven by internal energy. 

\par
\begin{figure}
\centering
\includegraphics[scale=0.4]{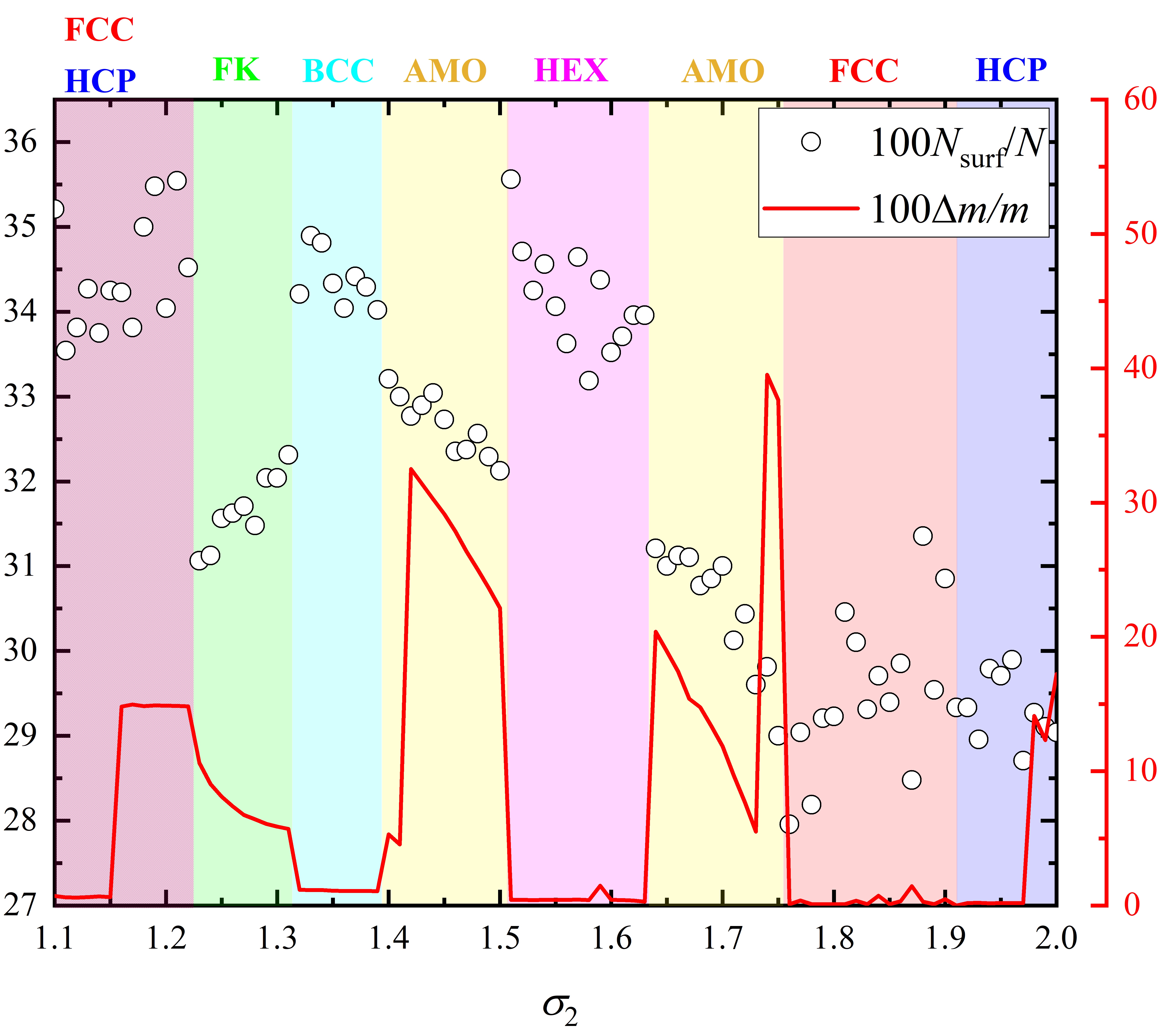}
\caption{\ojo{Percentage of monomers lying on the external surface of the formed cluster, $100 N_{surf}/N$ (left $y$-axis, scattered points) and percentage difference in the number of neighbors $100 \Delta m / m$ (right $y$-axis, red line) as a function of $\sigma_2$. $\Delta m$ is defined as the number of neighbors as predicted by the geometric model, $m$, minus the number of neighbors as calculated in the computer-generated system configurations. Background colors correspond to the dominant (crystal, F-K or amorphous) phase as observed in the MC simulations.}} 
\label{surfaceplot}
\end{figure}

 \ojo{The left panel of Fig. \ref{surfaceplot} (scattered open symbols) shows the percentage of the monomers lying on the external surface of the formed cluster as a function of interaction range, $\sigma_2$. Quite small deviations occur within the whole range, with around  two thirds of the sites having a fully developed Voronoi cell. Between the formed crystals the HEX one shows the highest surface to volume ratio. In general, the more compact and the more spherical the cluster the fewer the sites on the surface. Two distinct trends can be further observed: the number of surface atoms increases as a function of $\sigma_2$ for the Frank-Kasper phase but decreases monotonically in the amorphous zones. The right panel of Fig. \ref{surfaceplot} (red solid line) shows the percentage difference of the number of neighbors, $100 \Delta m / m$, on $\sigma_2$. $\Delta m$ is defined here as a the number of neighbors, as predicted by the proposed geometric model, minus the number of neighbors as calculated in the computer-generated system configurations including all snapshots in the equilibrated part of the MC trajectory. Very good to excellent agreement is observed in all regions where the crystal formed in the MC simulations coincides with the dominant one as predicted by the geometric model. Large deviations correspond to the two amorphous zones, an expected trend as these regions are bare of any crystal order. These are further accompanied by abrupt changes marking the AMO $\Leftrightarrow$ HEX and HEX $\Leftrightarrow$ AMO transitions. In the early regime a sharp discrepancy is observed as the expected BCC crystal, according to the model, is not encountered in the computer simulations which lead instead to rHCP morphologies of mixed FCC and HCP character. In parallel, the difference in the number of neighbors adopts the lowest values in the HEX- and FCC-dominated regions. It should be further noted that for the whole interaction range studied here no negative values in $\Delta m$ are observed. If such values existed it would mean that the MC simulations generate a stable crystal which is denser and thus different than the reference ones incorporated in the geometric neighbor model.}

\section{\label{sec:level4}Conclusions\protect\\}

\par The \textit{Digital Alchemy} proposed by Anders \textit{et al.} \cite{RN1814} and recent important advances in the synthesis and characterization show how the use of building blocks of low dimensionality by controlling chain stiffness and molecular architecture \cite{RN1875, RN1877, RN1880, RN1874} can lead to the design of tailored colloidal polymers and molecules. The present work demonstrates, at a fundamental level and utilizing a highly idealized model, how fine tuning a single interaction parameter can be used to obtain a rich assortment of target crystal structures in bulk 3-D and ultra-thin 2-D systems of hard colloidal chains of attractive monomers. \ojoo{Towards this, first, we propose a simple geometric model, based on the cumulative number of neighbors, to predict the dominant crystal as a function of attraction range. Then, we embark on Monte Carlo simulations, using the square well potential to equilibrate and successively identify the computer-generated polymer clusters. The flatness of the square well potential makes it ideal to compare the computer-generated structures against the reference crystals predicted by the geometric model. By tuning the attraction range a wealth of well defined ordered structures is observed including hexagonal closed packed, face centered cubic, simple hexagonal and body centered cubic crystals in 3D and triangular and cubic crystals in 2D. The 3D ordered morphologies are further accompanied by Frank-Kasper phases of the $\sigma$ variant. Interestingly enough the expected transitions between the HEX and BCC crystals are suppressed by the presence of amorphous zones where no traces of crystallization can be detected. Albeit its simplicity the proposed geometrical model, as demonstrated by the Monte Carlo simulations, is able to predict the dominant phases and the corresponding transitions with surprising accuracy, especially in 2D.} 
\par Current efforts focus on gauging the corresponding phase behavior of semi-flexible polymer chains under the same simulation conditions in the bulk (3D) and in extremely confined thin films (2D). 

\section{\label{sec:level5}Acknowledgements\protect\\}

\par M.H. deeply appreciates the kind hospitality of the COMSE group (NTUA, Athens, Greece) during his stay. Authors acknowledge support through projects PID2021-127533NB-I00 and RTI2018-097338-B-I00 of MICINN/FEDER (Ministerio de Ciencia e Innovación, Fondo Europeo de Desarrollo Regional). M.H. and D.M.F. acknowledge financial support through Programa Propio UPM Santander of Universidad Politécnica de Madrid (UPM) and Santander Bank. The authors gratefully acknowledge UPM for providing computing resources on the Magerit supercomputer through projects r553, r727, s341, t736 and u242.

\bibliography{Main_bib_LaTeX.bib}

\begin{thebibliography}{81}%
\makeatletter
\providecommand \@ifxundefined [1]{%
 \@ifx{#1\undefined}
}%
\providecommand \@ifnum [1]{%
 \ifnum #1\expandafter \@firstoftwo
 \else \expandafter \@secondoftwo
 \fi
}%
\providecommand \@ifx [1]{%
 \ifx #1\expandafter \@firstoftwo
 \else \expandafter \@secondoftwo
 \fi
}%
\providecommand \natexlab [1]{#1}%
\providecommand \enquote  [1]{``#1''}%
\providecommand \bibnamefont  [1]{#1}%
\providecommand \bibfnamefont [1]{#1}%
\providecommand \citenamefont [1]{#1}%
\providecommand \href@noop [0]{\@secondoftwo}%
\providecommand \href [0]{\begingroup \@sanitize@url \@href}%
\providecommand \@href[1]{\@@startlink{#1}\@@href}%
\providecommand \@@href[1]{\endgroup#1\@@endlink}%
\providecommand \@sanitize@url [0]{\catcode `\\12\catcode `\$12\catcode
  `\&12\catcode `\#12\catcode `\^12\catcode `\_12\catcode `\%12\relax}%
\providecommand \@@startlink[1]{}%
\providecommand \@@endlink[0]{}%
\providecommand \url  [0]{\begingroup\@sanitize@url \@url }%
\providecommand \@url [1]{\endgroup\@href {#1}{\urlprefix }}%
\providecommand \urlprefix  [0]{URL }%
\providecommand \Eprint [0]{\href }%
\providecommand \doibase [0]{https://doi.org/}%
\providecommand \selectlanguage [0]{\@gobble}%
\providecommand \bibinfo  [0]{\@secondoftwo}%
\providecommand \bibfield  [0]{\@secondoftwo}%
\providecommand \translation [1]{[#1]}%
\providecommand \BibitemOpen [0]{}%
\providecommand \bibitemStop [0]{}%
\providecommand \bibitemNoStop [0]{.\EOS\space}%
\providecommand \EOS [0]{\spacefactor3000\relax}%
\providecommand \BibitemShut  [1]{\csname bibitem#1\endcsname}%
\let\auto@bib@innerbib\@empty
%</preamble>
\bibitem [{\citenamefont {van Anders}\ \emph {et~al.}(2015)\citenamefont {van
  Anders}, \citenamefont {Klotsa}, \citenamefont {Karas}, \citenamefont
  {Dodd},\ and\ \citenamefont {Glotzer}}]{RN1814}%
  \BibitemOpen
  \bibfield  {author} {\bibinfo {author} {\bibfnamefont {G.}~\bibnamefont {van
  Anders}}, \bibinfo {author} {\bibfnamefont {D.}~\bibnamefont {Klotsa}},
  \bibinfo {author} {\bibfnamefont {A.~S.}\ \bibnamefont {Karas}}, \bibinfo
  {author} {\bibfnamefont {P.~M.}\ \bibnamefont {Dodd}},\ and\ \bibinfo
  {author} {\bibfnamefont {S.~C.}\ \bibnamefont {Glotzer}},\ }\bibfield
  {title} {\bibinfo {title} {Digital alchemy for materials design: Colloids and
  beyond},\ }\href {https://doi.org/10.1021/acsnano.5b04181} {\bibfield
  {journal} {\bibinfo  {journal} {Acs Nano}\ }\textbf {\bibinfo {volume} {9}},\
  \bibinfo {pages} {9542} (\bibinfo {year} {2015})}\BibitemShut {NoStop}%
\bibitem [{\citenamefont {Li}\ \emph {et~al.}(2021)\citenamefont {Li},
  \citenamefont {Zhang}, \citenamefont {Andre},\ and\ \citenamefont
  {Chen}}]{RN1895}%
  \BibitemOpen
  \bibfield  {author} {\bibinfo {author} {\bibfnamefont {B.}~\bibnamefont
  {Li}}, \bibinfo {author} {\bibfnamefont {S.}~\bibnamefont {Zhang}}, \bibinfo
  {author} {\bibfnamefont {J.~S.}\ \bibnamefont {Andre}},\ and\ \bibinfo
  {author} {\bibfnamefont {Z.}~\bibnamefont {Chen}},\ }\bibfield  {title}
  {\bibinfo {title} {Relaxation behavior of polymer thin films: Effects of free
  surface, buried interface, and geometrical confinement},\ }\bibfield
  {journal} {\bibinfo  {journal} {Progress in Polymer Science}\ }\textbf
  {\bibinfo {volume} {120}},\ \href
  {https://doi.org/10.1016/j.progpolymsci.2021.101431}
  {10.1016/j.progpolymsci.2021.101431} (\bibinfo {year} {2021})\BibitemShut
  {NoStop}%
\bibitem [{\citenamefont {Damasceno}\ \emph {et~al.}(2012)\citenamefont
  {Damasceno}, \citenamefont {Engel},\ and\ \citenamefont {Glotzer}}]{RN116}%
  \BibitemOpen
  \bibfield  {author} {\bibinfo {author} {\bibfnamefont {P.~F.}\ \bibnamefont
  {Damasceno}}, \bibinfo {author} {\bibfnamefont {M.}~\bibnamefont {Engel}},\
  and\ \bibinfo {author} {\bibfnamefont {S.~C.}\ \bibnamefont {Glotzer}},\
  }\bibfield  {title} {\bibinfo {title} {Predictive self-assembly of polyhedra
  into complex structures},\ }\href {https://doi.org/10.1126/science.1220869}
  {\bibfield  {journal} {\bibinfo  {journal} {Science}\ }\textbf {\bibinfo
  {volume} {337}},\ \bibinfo {pages} {453} (\bibinfo {year}
  {2012})}\BibitemShut {NoStop}%
\bibitem [{\citenamefont {Stillinger}\ \emph {et~al.}(1964)\citenamefont
  {Stillinger}, \citenamefont {Kornegay},\ and\ \citenamefont
  {Dimarzio}}]{RN1896}%
  \BibitemOpen
  \bibfield  {author} {\bibinfo {author} {\bibfnamefont {F.~H.}\ \bibnamefont
  {Stillinger}}, \bibinfo {author} {\bibfnamefont {R.~L.}\ \bibnamefont
  {Kornegay}},\ and\ \bibinfo {author} {\bibfnamefont {E.~A.}\ \bibnamefont
  {Dimarzio}},\ }\bibfield  {title} {\bibinfo {title} {Systematic approach to
  explanation of rigid disk phase transition},\ }\href
  {https://doi.org/10.1063/1.1725362} {\bibfield  {journal} {\bibinfo
  {journal} {Journal of Chemical Physics}\ }\textbf {\bibinfo {volume} {40}},\
  \bibinfo {pages} {1564} (\bibinfo {year} {1964})}\BibitemShut {NoStop}%
\bibitem [{\citenamefont {Hinrichsen}\ \emph {et~al.}(1990)\citenamefont
  {Hinrichsen}, \citenamefont {Feder},\ and\ \citenamefont {Jossang}}]{RN1898}%
  \BibitemOpen
  \bibfield  {author} {\bibinfo {author} {\bibfnamefont {E.~L.}\ \bibnamefont
  {Hinrichsen}}, \bibinfo {author} {\bibfnamefont {J.}~\bibnamefont {Feder}},\
  and\ \bibinfo {author} {\bibfnamefont {T.}~\bibnamefont {Jossang}},\
  }\bibfield  {title} {\bibinfo {title} {Random packing of disks in 2
  dimensions},\ }\href {https://doi.org/10.1103/PhysRevA.41.4199} {\bibfield
  {journal} {\bibinfo  {journal} {Physical Review A}\ }\textbf {\bibinfo
  {volume} {41}},\ \bibinfo {pages} {4199} (\bibinfo {year}
  {1990})}\BibitemShut {NoStop}%
\bibitem [{\citenamefont {Visscher}\ and\ \citenamefont
  {Bolsterl.M}(1972)}]{RN1897}%
  \BibitemOpen
  \bibfield  {author} {\bibinfo {author} {\bibfnamefont {W.~M.}\ \bibnamefont
  {Visscher}}\ and\ \bibinfo {author} {\bibnamefont {Bolsterl.M}},\ }\bibfield
  {title} {\bibinfo {title} {Random packing of equal and unequal spheres in 2
  and 3 dimensions},\ }\href {https://doi.org/10.1038/239504a0} {\bibfield
  {journal} {\bibinfo  {journal} {Nature}\ }\textbf {\bibinfo {volume} {239}},\
  \bibinfo {pages} {504} (\bibinfo {year} {1972})}\BibitemShut {NoStop}%
\bibitem [{\citenamefont {Meyer}\ \emph {et~al.}(2010)\citenamefont {Meyer},
  \citenamefont {Song}, \citenamefont {Jin}, \citenamefont {Wang},\ and\
  \citenamefont {Makse}}]{RN1899}%
  \BibitemOpen
  \bibfield  {author} {\bibinfo {author} {\bibfnamefont {S.}~\bibnamefont
  {Meyer}}, \bibinfo {author} {\bibfnamefont {C.~M.}\ \bibnamefont {Song}},
  \bibinfo {author} {\bibfnamefont {Y.~L.}\ \bibnamefont {Jin}}, \bibinfo
  {author} {\bibfnamefont {K.}~\bibnamefont {Wang}},\ and\ \bibinfo {author}
  {\bibfnamefont {H.~A.}\ \bibnamefont {Makse}},\ }\bibfield  {title} {\bibinfo
  {title} {Jamming in two-dimensional packings},\ }\href
  {https://doi.org/10.1016/j.physa.2010.07.030} {\bibfield  {journal} {\bibinfo
   {journal} {Physica a-Statistical Mechanics and Its Applications}\ }\textbf
  {\bibinfo {volume} {389}},\ \bibinfo {pages} {5137} (\bibinfo {year}
  {2010})}\BibitemShut {NoStop}%
\bibitem [{\citenamefont {Gong}\ \emph {et~al.}(2017)\citenamefont {Gong},
  \citenamefont {Hueckel}, \citenamefont {Yi},\ and\ \citenamefont
  {Sacanna}}]{RN1819}%
  \BibitemOpen
  \bibfield  {author} {\bibinfo {author} {\bibfnamefont {Z.}~\bibnamefont
  {Gong}}, \bibinfo {author} {\bibfnamefont {T.}~\bibnamefont {Hueckel}},
  \bibinfo {author} {\bibfnamefont {G.~R.}\ \bibnamefont {Yi}},\ and\ \bibinfo
  {author} {\bibfnamefont {S.}~\bibnamefont {Sacanna}},\ }\bibfield  {title}
  {\bibinfo {title} {Patchy particles made by colloidal fusion},\ }\href
  {https://doi.org/10.1038/nature23901} {\bibfield  {journal} {\bibinfo
  {journal} {Nature}\ }\textbf {\bibinfo {volume} {550}},\ \bibinfo {pages}
  {234} (\bibinfo {year} {2017})}\BibitemShut {NoStop}%
\bibitem [{\citenamefont {Kennedy}\ \emph {et~al.}(2022)\citenamefont
  {Kennedy}, \citenamefont {Sayasilpi}, \citenamefont {Schall},\ and\
  \citenamefont {Meijer}}]{RN1871}%
  \BibitemOpen
  \bibfield  {author} {\bibinfo {author} {\bibfnamefont {C.~L.}\ \bibnamefont
  {Kennedy}}, \bibinfo {author} {\bibfnamefont {D.}~\bibnamefont {Sayasilpi}},
  \bibinfo {author} {\bibfnamefont {P.}~\bibnamefont {Schall}},\ and\ \bibinfo
  {author} {\bibfnamefont {J.~M.}\ \bibnamefont {Meijer}},\ }\bibfield  {title}
  {\bibinfo {title} {Self-assembly of colloidal cube superstructures with
  critical casimir attractions},\ }\bibfield  {journal} {\bibinfo  {journal}
  {Journal of Physics-Condensed Matter}\ }\textbf {\bibinfo {volume} {34}},\
  \href {https://doi.org/10.1088/1361-648X/ac5866} {10.1088/1361-648X/ac5866}
  (\bibinfo {year} {2022})\BibitemShut {NoStop}%
\bibitem [{\citenamefont {Kuijk}\ \emph {et~al.}(2011)\citenamefont {Kuijk},
  \citenamefont {van Blaaderen},\ and\ \citenamefont {Imhof}}]{RN1872}%
  \BibitemOpen
  \bibfield  {author} {\bibinfo {author} {\bibfnamefont {A.}~\bibnamefont
  {Kuijk}}, \bibinfo {author} {\bibfnamefont {A.}~\bibnamefont {van
  Blaaderen}},\ and\ \bibinfo {author} {\bibfnamefont {A.}~\bibnamefont
  {Imhof}},\ }\bibfield  {title} {\bibinfo {title} {Synthesis of monodisperse,
  rodlike silica colloids with tunable aspect ratio},\ }\href
  {https://doi.org/10.1021/ja109524h} {\bibfield  {journal} {\bibinfo
  {journal} {Journal of the American Chemical Society}\ }\textbf {\bibinfo
  {volume} {133}},\ \bibinfo {pages} {2346} (\bibinfo {year}
  {2011})}\BibitemShut {NoStop}%
\bibitem [{\citenamefont {Sacanna}\ \emph {et~al.}(2010)\citenamefont
  {Sacanna}, \citenamefont {Irvine}, \citenamefont {Chaikin},\ and\
  \citenamefont {Pine}}]{RN439}%
  \BibitemOpen
  \bibfield  {author} {\bibinfo {author} {\bibfnamefont {S.}~\bibnamefont
  {Sacanna}}, \bibinfo {author} {\bibfnamefont {W.~T.~M.}\ \bibnamefont
  {Irvine}}, \bibinfo {author} {\bibfnamefont {P.~M.}\ \bibnamefont
  {Chaikin}},\ and\ \bibinfo {author} {\bibfnamefont {D.~J.}\ \bibnamefont
  {Pine}},\ }\bibfield  {title} {\bibinfo {title} {Lock and key colloids},\
  }\href {https://doi.org/10.1038/nature08906} {\bibfield  {journal} {\bibinfo
  {journal} {Nature}\ }\textbf {\bibinfo {volume} {464}},\ \bibinfo {pages}
  {575} (\bibinfo {year} {2010})}\BibitemShut {NoStop}%
\bibitem [{\citenamefont {Sacanna}\ \emph {et~al.}(2013)\citenamefont
  {Sacanna}, \citenamefont {Korpics}, \citenamefont {Rodriguez}, \citenamefont
  {Colon-Melendez}, \citenamefont {Kim}, \citenamefont {Pine},\ and\
  \citenamefont {Yi}}]{RN1818}%
  \BibitemOpen
  \bibfield  {author} {\bibinfo {author} {\bibfnamefont {S.}~\bibnamefont
  {Sacanna}}, \bibinfo {author} {\bibfnamefont {M.}~\bibnamefont {Korpics}},
  \bibinfo {author} {\bibfnamefont {K.}~\bibnamefont {Rodriguez}}, \bibinfo
  {author} {\bibfnamefont {L.}~\bibnamefont {Colon-Melendez}}, \bibinfo
  {author} {\bibfnamefont {S.~H.}\ \bibnamefont {Kim}}, \bibinfo {author}
  {\bibfnamefont {D.~J.}\ \bibnamefont {Pine}},\ and\ \bibinfo {author}
  {\bibfnamefont {G.~R.}\ \bibnamefont {Yi}},\ }\bibfield  {title} {\bibinfo
  {title} {Shaping colloids for self-assembly},\ }\bibfield  {journal}
  {\bibinfo  {journal} {Nature Communications}\ }\textbf {\bibinfo {volume}
  {4}},\ \href {https://doi.org/10.1038/ncomms2694} {10.1038/ncomms2694}
  (\bibinfo {year} {2013})\BibitemShut {NoStop}%
\bibitem [{\citenamefont {Xia}\ \emph {et~al.}(2009)\citenamefont {Xia},
  \citenamefont {Xiong}, \citenamefont {Lim},\ and\ \citenamefont
  {Skrabalak}}]{RN1817}%
  \BibitemOpen
  \bibfield  {author} {\bibinfo {author} {\bibfnamefont {Y.~N.}\ \bibnamefont
  {Xia}}, \bibinfo {author} {\bibfnamefont {Y.~J.}\ \bibnamefont {Xiong}},
  \bibinfo {author} {\bibfnamefont {B.}~\bibnamefont {Lim}},\ and\ \bibinfo
  {author} {\bibfnamefont {S.~E.}\ \bibnamefont {Skrabalak}},\ }\bibfield
  {title} {\bibinfo {title} {Shape-controlled synthesis of metal nanocrystals:
  Simple chemistry meets complex physics?},\ }\href
  {https://doi.org/10.1002/anie.200802248} {\bibfield  {journal} {\bibinfo
  {journal} {Angewandte Chemie-International Edition}\ }\textbf {\bibinfo
  {volume} {48}},\ \bibinfo {pages} {60} (\bibinfo {year} {2009})}\BibitemShut
  {NoStop}%
\bibitem [{\citenamefont {Reiter}\ and\ \citenamefont {Sommer}(2008)}]{RN1462}%
  \BibitemOpen
  \bibfield  {author} {\bibinfo {author} {\bibfnamefont {G.}~\bibnamefont
  {Reiter}}\ and\ \bibinfo {author} {\bibfnamefont {J.}~\bibnamefont
  {Sommer}},\ }\href {https://books.google.es/books?id=4wRqCQAAQBAJ} {\emph
  {\bibinfo {title} {Polymer Crystallization: Obervations, Concepts and
  Interpretations}}}\ (\bibinfo  {publisher} {Springer Berlin Heidelberg},\
  \bibinfo {year} {2008})\BibitemShut {NoStop}%
\bibitem [{\citenamefont {Fan}\ and\ \citenamefont {Walther}(2022)}]{RN1875}%
  \BibitemOpen
  \bibfield  {author} {\bibinfo {author} {\bibfnamefont {X.~L.}\ \bibnamefont
  {Fan}}\ and\ \bibinfo {author} {\bibfnamefont {A.}~\bibnamefont {Walther}},\
  }\bibfield  {title} {\bibinfo {title} {1d colloidal chains: recent progress
  from formation to emergent properties and applications},\ }\href
  {https://doi.org/10.1039/d2cs00112h} {\bibfield  {journal} {\bibinfo
  {journal} {Chemical Society Reviews}\ }\textbf {\bibinfo {volume} {51}},\
  \bibinfo {pages} {4023} (\bibinfo {year} {2022})}\BibitemShut {NoStop}%
\bibitem [{\citenamefont {Li}\ \emph {et~al.}(2011)\citenamefont {Li},
  \citenamefont {Josephson},\ and\ \citenamefont {Stein}}]{RN1873}%
  \BibitemOpen
  \bibfield  {author} {\bibinfo {author} {\bibfnamefont {F.}~\bibnamefont
  {Li}}, \bibinfo {author} {\bibfnamefont {D.~P.}\ \bibnamefont {Josephson}},\
  and\ \bibinfo {author} {\bibfnamefont {A.}~\bibnamefont {Stein}},\ }\bibfield
   {title} {\bibinfo {title} {Colloidal assembly: The road from particles to
  colloidal molecules and crystals},\ }\href
  {https://doi.org/10.1002/anie.201001451} {\bibfield  {journal} {\bibinfo
  {journal} {Angewandte Chemie-International Edition}\ }\textbf {\bibinfo
  {volume} {50}},\ \bibinfo {pages} {360} (\bibinfo {year} {2011})}\BibitemShut
  {NoStop}%
\bibitem [{\citenamefont {Chakraborty}\ \emph {et~al.}(2022)\citenamefont
  {Chakraborty}, \citenamefont {Pearce}, \citenamefont {Verweij}, \citenamefont
  {Matysik}, \citenamefont {Giomi},\ and\ \citenamefont {Kraft}}]{RN1877}%
  \BibitemOpen
  \bibfield  {author} {\bibinfo {author} {\bibfnamefont {I.}~\bibnamefont
  {Chakraborty}}, \bibinfo {author} {\bibfnamefont {D.~J.~G.}\ \bibnamefont
  {Pearce}}, \bibinfo {author} {\bibfnamefont {R.~W.}\ \bibnamefont {Verweij}},
  \bibinfo {author} {\bibfnamefont {S.~C.}\ \bibnamefont {Matysik}}, \bibinfo
  {author} {\bibfnamefont {L.}~\bibnamefont {Giomi}},\ and\ \bibinfo {author}
  {\bibfnamefont {D.~J.}\ \bibnamefont {Kraft}},\ }\bibfield  {title} {\bibinfo
  {title} {Self-assembly dynamics of reconfigurable colloidal molecules},\
  }\href {https://doi.org/10.1021/acsnano.1c09088} {\bibfield  {journal}
  {\bibinfo  {journal} {Acs Nano}\ }\textbf {\bibinfo {volume} {16}},\ \bibinfo
  {pages} {2471} (\bibinfo {year} {2022})}\BibitemShut {NoStop}%
\bibitem [{\citenamefont {Huil}\ \emph {et~al.}(2015)\citenamefont {Huil},
  \citenamefont {Pinna}, \citenamefont {Char},\ and\ \citenamefont
  {Pyun}}]{RN1879}%
  \BibitemOpen
  \bibfield  {author} {\bibinfo {author} {\bibfnamefont {L.~J.}\ \bibnamefont
  {Huil}}, \bibinfo {author} {\bibfnamefont {N.}~\bibnamefont {Pinna}},
  \bibinfo {author} {\bibfnamefont {K.}~\bibnamefont {Char}},\ and\ \bibinfo
  {author} {\bibfnamefont {J.}~\bibnamefont {Pyun}},\ }\bibfield  {title}
  {\bibinfo {title} {Colloidal polymers from inorganic nanoparticle monomers},\
  }\href {https://doi.org/10.1016/j.progpolymsci.2014.08.003} {\bibfield
  {journal} {\bibinfo  {journal} {Progress in Polymer Science}\ }\textbf
  {\bibinfo {volume} {40}},\ \bibinfo {pages} {85} (\bibinfo {year}
  {2015})}\BibitemShut {NoStop}%
\bibitem [{\citenamefont {Li}\ \emph {et~al.}(2020)\citenamefont {Li},
  \citenamefont {Palis}, \citenamefont {Merindol}, \citenamefont {Majimel},
  \citenamefont {Ravaine},\ and\ \citenamefont {Duguet}}]{RN1880}%
  \BibitemOpen
  \bibfield  {author} {\bibinfo {author} {\bibfnamefont {W.~Y.}\ \bibnamefont
  {Li}}, \bibinfo {author} {\bibfnamefont {H.}~\bibnamefont {Palis}}, \bibinfo
  {author} {\bibfnamefont {R.}~\bibnamefont {Merindol}}, \bibinfo {author}
  {\bibfnamefont {J.}~\bibnamefont {Majimel}}, \bibinfo {author} {\bibfnamefont
  {S.}~\bibnamefont {Ravaine}},\ and\ \bibinfo {author} {\bibfnamefont
  {E.}~\bibnamefont {Duguet}},\ }\bibfield  {title} {\bibinfo {title}
  {Colloidal molecules and patchy particles: complementary concepts, synthesis
  and self-assembly},\ }\href {https://doi.org/10.1039/c9cs00804g} {\bibfield
  {journal} {\bibinfo  {journal} {Chemical Society Reviews}\ }\textbf {\bibinfo
  {volume} {49}},\ \bibinfo {pages} {1955} (\bibinfo {year}
  {2020})}\BibitemShut {NoStop}%
\bibitem [{\citenamefont {Martinez-Pedrero}\ \emph {et~al.}(2021)\citenamefont
  {Martinez-Pedrero}, \citenamefont {Gonzalez-Banciella}, \citenamefont
  {Camino}, \citenamefont {Mateos-Maroto}, \citenamefont {Ortega},
  \citenamefont {Rubio}, \citenamefont {Pagonabarraga},\ and\ \citenamefont
  {Calero}}]{RN1874}%
  \BibitemOpen
  \bibfield  {author} {\bibinfo {author} {\bibfnamefont {F.}~\bibnamefont
  {Martinez-Pedrero}}, \bibinfo {author} {\bibfnamefont {A.}~\bibnamefont
  {Gonzalez-Banciella}}, \bibinfo {author} {\bibfnamefont {A.}~\bibnamefont
  {Camino}}, \bibinfo {author} {\bibfnamefont {A.}~\bibnamefont
  {Mateos-Maroto}}, \bibinfo {author} {\bibfnamefont {F.}~\bibnamefont
  {Ortega}}, \bibinfo {author} {\bibfnamefont {R.~G.}\ \bibnamefont {Rubio}},
  \bibinfo {author} {\bibfnamefont {I.}~\bibnamefont {Pagonabarraga}},\ and\
  \bibinfo {author} {\bibfnamefont {C.}~\bibnamefont {Calero}},\ }\bibfield
  {title} {\bibinfo {title} {Static and dynamic self-assembly of
  pearl-like-chains of magnetic colloids confined at fluid interfaces},\
  }\bibfield  {journal} {\bibinfo  {journal} {Small}\ }\textbf {\bibinfo
  {volume} {17}},\ \href {https://doi.org/10.1002/smll.202101188}
  {10.1002/smll.202101188} (\bibinfo {year} {2021})\BibitemShut {NoStop}%
\bibitem [{\citenamefont {Stuij}\ \emph {et~al.}(2019)\citenamefont {Stuij},
  \citenamefont {van Doorn}, \citenamefont {Kodger}, \citenamefont {Sprakel},
  \citenamefont {Coulais},\ and\ \citenamefont {Schall}}]{RN1878}%
  \BibitemOpen
  \bibfield  {author} {\bibinfo {author} {\bibfnamefont {S.}~\bibnamefont
  {Stuij}}, \bibinfo {author} {\bibfnamefont {J.~M.}\ \bibnamefont {van
  Doorn}}, \bibinfo {author} {\bibfnamefont {T.}~\bibnamefont {Kodger}},
  \bibinfo {author} {\bibfnamefont {J.}~\bibnamefont {Sprakel}}, \bibinfo
  {author} {\bibfnamefont {C.}~\bibnamefont {Coulais}},\ and\ \bibinfo {author}
  {\bibfnamefont {P.}~\bibnamefont {Schall}},\ }\bibfield  {title} {\bibinfo
  {title} {Stochastic buckling of self-assembled colloidal structures},\
  }\bibfield  {journal} {\bibinfo  {journal} {Physical Review Research}\
  }\textbf {\bibinfo {volume} {1}},\ \href
  {https://doi.org/10.1103/PhysRevResearch.1.023033}
  {10.1103/PhysRevResearch.1.023033} (\bibinfo {year} {2019})\BibitemShut
  {NoStop}%
\bibitem [{\citenamefont {Verweij}\ \emph {et~al.}(2020)\citenamefont
  {Verweij}, \citenamefont {Moerman}, \citenamefont {Ligthart}, \citenamefont
  {Huijnen}, \citenamefont {Groenewold}, \citenamefont {Kegel}, \citenamefont
  {van Blaaderen},\ and\ \citenamefont {Kraft}}]{RN1876}%
  \BibitemOpen
  \bibfield  {author} {\bibinfo {author} {\bibfnamefont {R.~W.}\ \bibnamefont
  {Verweij}}, \bibinfo {author} {\bibfnamefont {P.~G.}\ \bibnamefont
  {Moerman}}, \bibinfo {author} {\bibfnamefont {N.~E.~G.}\ \bibnamefont
  {Ligthart}}, \bibinfo {author} {\bibfnamefont {L.~P.~P.}\ \bibnamefont
  {Huijnen}}, \bibinfo {author} {\bibfnamefont {J.}~\bibnamefont {Groenewold}},
  \bibinfo {author} {\bibfnamefont {W.~K.}\ \bibnamefont {Kegel}}, \bibinfo
  {author} {\bibfnamefont {A.}~\bibnamefont {van Blaaderen}},\ and\ \bibinfo
  {author} {\bibfnamefont {D.~J.}\ \bibnamefont {Kraft}},\ }\bibfield  {title}
  {\bibinfo {title} {Flexibility-induced effects in the brownian motion of
  colloidal trimers},\ }\bibfield  {journal} {\bibinfo  {journal} {Physical
  Review Research}\ }\textbf {\bibinfo {volume} {2}},\ \href
  {https://doi.org/10.1103/PhysRevResearch.2.033136}
  {10.1103/PhysRevResearch.2.033136} (\bibinfo {year} {2020})\BibitemShut
  {NoStop}%
\bibitem [{\citenamefont {Vutukuri}\ \emph {et~al.}(2012)\citenamefont
  {Vutukuri}, \citenamefont {Demirors}, \citenamefont {Peng}, \citenamefont
  {van Oostrum}, \citenamefont {Imhof},\ and\ \citenamefont {van
  Blaaderen}}]{RN377}%
  \BibitemOpen
  \bibfield  {author} {\bibinfo {author} {\bibfnamefont {H.~R.}\ \bibnamefont
  {Vutukuri}}, \bibinfo {author} {\bibfnamefont {A.~F.}\ \bibnamefont
  {Demirors}}, \bibinfo {author} {\bibfnamefont {B.}~\bibnamefont {Peng}},
  \bibinfo {author} {\bibfnamefont {P.~D.~J.}\ \bibnamefont {van Oostrum}},
  \bibinfo {author} {\bibfnamefont {A.}~\bibnamefont {Imhof}},\ and\ \bibinfo
  {author} {\bibfnamefont {A.}~\bibnamefont {van Blaaderen}},\ }\bibfield
  {title} {\bibinfo {title} {Colloidal analogues of charged and uncharged
  polymer chains with tunable stiffness},\ }\href
  {https://doi.org/10.1002/anie.201202592} {\bibfield  {journal} {\bibinfo
  {journal} {Angewandte Chemie-International Edition}\ }\textbf {\bibinfo
  {volume} {51}},\ \bibinfo {pages} {11249} (\bibinfo {year}
  {2012})}\BibitemShut {NoStop}%
\bibitem [{\citenamefont {Dietz}\ and\ \citenamefont {Hoy}(2020)}]{RN1558}%
  \BibitemOpen
  \bibfield  {author} {\bibinfo {author} {\bibfnamefont {J.~D.}\ \bibnamefont
  {Dietz}}\ and\ \bibinfo {author} {\bibfnamefont {R.~S.}\ \bibnamefont
  {Hoy}},\ }\bibfield  {title} {\bibinfo {title} {Two-stage athermal
  solidification of semiflexible polymers and fibers},\ }\href
  {https://doi.org/10.1039/d0sm00754d} {\bibfield  {journal} {\bibinfo
  {journal} {Soft Matter}\ }\textbf {\bibinfo {volume} {16}},\ \bibinfo {pages}
  {6206} (\bibinfo {year} {2020})}\BibitemShut {NoStop}%
\bibitem [{\citenamefont {Mahynski}\ \emph {et~al.}(2015)\citenamefont
  {Mahynski}, \citenamefont {Kumar},\ and\ \citenamefont
  {Panagiotopoulos}}]{RN1748}%
  \BibitemOpen
  \bibfield  {author} {\bibinfo {author} {\bibfnamefont {N.~A.}\ \bibnamefont
  {Mahynski}}, \bibinfo {author} {\bibfnamefont {S.~K.}\ \bibnamefont
  {Kumar}},\ and\ \bibinfo {author} {\bibfnamefont {A.~Z.}\ \bibnamefont
  {Panagiotopoulos}},\ }\bibfield  {title} {\bibinfo {title} {Relative
  stability of the fcc and hcp polymorphs with interacting polymers},\ }\href
  {https://doi.org/10.1039/c4sm02191f} {\bibfield  {journal} {\bibinfo
  {journal} {Soft Matter}\ }\textbf {\bibinfo {volume} {11}},\ \bibinfo {pages}
  {280} (\bibinfo {year} {2015})}\BibitemShut {NoStop}%
\bibitem [{\citenamefont {Mahynski}\ \emph {et~al.}(2014)\citenamefont
  {Mahynski}, \citenamefont {Panagiotopoulos}, \citenamefont {Meng},\ and\
  \citenamefont {Kumar}}]{RN1726}%
  \BibitemOpen
  \bibfield  {author} {\bibinfo {author} {\bibfnamefont {N.~A.}\ \bibnamefont
  {Mahynski}}, \bibinfo {author} {\bibfnamefont {A.~Z.}\ \bibnamefont
  {Panagiotopoulos}}, \bibinfo {author} {\bibfnamefont {D.}~\bibnamefont
  {Meng}},\ and\ \bibinfo {author} {\bibfnamefont {S.~K.}\ \bibnamefont
  {Kumar}},\ }\bibfield  {title} {\bibinfo {title} {Stabilizing colloidal
  crystals by leveraging void distributions},\ }\bibfield  {journal} {\bibinfo
  {journal} {Nature Communications}\ }\textbf {\bibinfo {volume} {5}},\ \href
  {https://doi.org/10.1038/ncomms5472} {10.1038/ncomms5472} (\bibinfo {year}
  {2014})\BibitemShut {NoStop}%
\bibitem [{\citenamefont {Shakirov}(2019)}]{RN1339}%
  \BibitemOpen
  \bibfield  {author} {\bibinfo {author} {\bibfnamefont {T.}~\bibnamefont
  {Shakirov}},\ }\bibfield  {title} {\bibinfo {title} {Crystallisation in melts
  of short, semi-flexible hard-sphere polymer chains: The role of the
  non-bonded interaction range},\ }\bibfield  {journal} {\bibinfo  {journal}
  {Entropy}\ }\textbf {\bibinfo {volume} {21}},\ \href
  {https://doi.org/10.3390/e21090856} {10.3390/e21090856} (\bibinfo {year}
  {2019})\BibitemShut {NoStop}%
\bibitem [{\citenamefont {Verweij}\ \emph {et~al.}(2021)\citenamefont
  {Verweij}, \citenamefont {Moerman}, \citenamefont {Huijnen}, \citenamefont
  {Ligthart}, \citenamefont {Chakraborty}, \citenamefont {Groenewold},
  \citenamefont {Kegel}, \citenamefont {van Blaaderen},\ and\ \citenamefont
  {Kraft}}]{RN1809}%
  \BibitemOpen
  \bibfield  {author} {\bibinfo {author} {\bibfnamefont {R.~W.}\ \bibnamefont
  {Verweij}}, \bibinfo {author} {\bibfnamefont {P.~G.}\ \bibnamefont
  {Moerman}}, \bibinfo {author} {\bibfnamefont {L.~P.~P.}\ \bibnamefont
  {Huijnen}}, \bibinfo {author} {\bibfnamefont {N.~E.~G.}\ \bibnamefont
  {Ligthart}}, \bibinfo {author} {\bibfnamefont {I.}~\bibnamefont
  {Chakraborty}}, \bibinfo {author} {\bibfnamefont {J.}~\bibnamefont
  {Groenewold}}, \bibinfo {author} {\bibfnamefont {W.~K.}\ \bibnamefont
  {Kegel}}, \bibinfo {author} {\bibfnamefont {A.}~\bibnamefont {van
  Blaaderen}},\ and\ \bibinfo {author} {\bibfnamefont {D.~J.}\ \bibnamefont
  {Kraft}},\ }\bibfield  {title} {\bibinfo {title} {Conformations and diffusion
  of flexibly linked colloidal chains},\ }\bibfield  {journal} {\bibinfo
  {journal} {Journal of Physics-Materials}\ }\textbf {\bibinfo {volume} {4}},\
  \href {https://doi.org/10.1088/2515-7639/abf571} {10.1088/2515-7639/abf571}
  (\bibinfo {year} {2021})\BibitemShut {NoStop}%
\bibitem [{\citenamefont {Karayiannis}\ \emph {et~al.}(2015)\citenamefont
  {Karayiannis}, \citenamefont {Foteinopoulou},\ and\ \citenamefont
  {Laso}}]{RN497}%
  \BibitemOpen
  \bibfield  {author} {\bibinfo {author} {\bibfnamefont {N.~C.}\ \bibnamefont
  {Karayiannis}}, \bibinfo {author} {\bibfnamefont {K.}~\bibnamefont
  {Foteinopoulou}},\ and\ \bibinfo {author} {\bibfnamefont {M.}~\bibnamefont
  {Laso}},\ }\bibfield  {title} {\bibinfo {title} {The role of bond tangency
  and bond gap in hard sphere crystallization of chains},\ }\href
  {https://doi.org/10.1039/c4sm02707h} {\bibfield  {journal} {\bibinfo
  {journal} {Soft Matter}\ }\textbf {\bibinfo {volume} {11}},\ \bibinfo {pages}
  {1688} (\bibinfo {year} {2015})}\BibitemShut {NoStop}%
\bibitem [{\citenamefont {Ni}\ and\ \citenamefont {Dijkstra}(2013)}]{RN319}%
  \BibitemOpen
  \bibfield  {author} {\bibinfo {author} {\bibfnamefont {R.}~\bibnamefont
  {Ni}}\ and\ \bibinfo {author} {\bibfnamefont {M.}~\bibnamefont {Dijkstra}},\
  }\bibfield  {title} {\bibinfo {title} {Effect of bond length fluctuations on
  crystal nucleation of hard bead chains},\ }\href
  {https://doi.org/10.1039/c2sm26969d} {\bibfield  {journal} {\bibinfo
  {journal} {Soft Matter}\ }\textbf {\bibinfo {volume} {9}},\ \bibinfo {pages}
  {365} (\bibinfo {year} {2013})}\BibitemShut {NoStop}%
\bibitem [{\citenamefont {Young}\ and\ \citenamefont {Alder}(1979)}]{RN1528}%
  \BibitemOpen
  \bibfield  {author} {\bibinfo {author} {\bibfnamefont {D.~A.}\ \bibnamefont
  {Young}}\ and\ \bibinfo {author} {\bibfnamefont {B.~J.}\ \bibnamefont
  {Alder}},\ }\bibfield  {title} {\bibinfo {title} {Studies in
  molecular-dynamics .17. phase-diagrams for step potentials in 2 and 3
  dimensions},\ }\href {https://doi.org/10.1063/1.437212} {\bibfield  {journal}
  {\bibinfo  {journal} {Journal of Chemical Physics}\ }\textbf {\bibinfo
  {volume} {70}},\ \bibinfo {pages} {473} (\bibinfo {year} {1979})}\BibitemShut
  {NoStop}%
\bibitem [{\citenamefont {Young}\ and\ \citenamefont {Alder}(1980)}]{RN1529}%
  \BibitemOpen
  \bibfield  {author} {\bibinfo {author} {\bibfnamefont {D.~A.}\ \bibnamefont
  {Young}}\ and\ \bibinfo {author} {\bibfnamefont {B.~J.}\ \bibnamefont
  {Alder}},\ }\bibfield  {title} {\bibinfo {title} {Studies in
  molecular-dynamics .18. the square-well phase-diagram},\ }\href
  {https://doi.org/10.1063/1.440393} {\bibfield  {journal} {\bibinfo  {journal}
  {Journal of Chemical Physics}\ }\textbf {\bibinfo {volume} {73}},\ \bibinfo
  {pages} {2430} (\bibinfo {year} {1980})}\BibitemShut {NoStop}%
\bibitem [{\citenamefont {Armas-Pérez}\ \emph {et~al.}(2014)\citenamefont
  {Armas-Pérez}, \citenamefont {Quintana-H}, \citenamefont {Chapela},
  \citenamefont {Velasco},\ and\ \citenamefont {Navascués}}]{RN1540}%
  \BibitemOpen
  \bibfield  {author} {\bibinfo {author} {\bibfnamefont {J.~C.}\ \bibnamefont
  {Armas-Pérez}}, \bibinfo {author} {\bibfnamefont {J.}~\bibnamefont
  {Quintana-H}}, \bibinfo {author} {\bibfnamefont {G.~A.}\ \bibnamefont
  {Chapela}}, \bibinfo {author} {\bibfnamefont {E.}~\bibnamefont {Velasco}},\
  and\ \bibinfo {author} {\bibfnamefont {G.}~\bibnamefont {Navascués}},\
  }\bibfield  {title} {\bibinfo {title} {Phase diagram of a square-well model
  in two dimensions},\ }\href {https://doi.org/10.1063/1.4863993} {\bibfield
  {journal} {\bibinfo  {journal} {The Journal of Chemical Physics}\ }\textbf
  {\bibinfo {volume} {140}},\ \bibinfo {pages} {064503} (\bibinfo {year}
  {2014})}\BibitemShut {NoStop}%
\bibitem [{\citenamefont {Attwood}\ and\ \citenamefont {Hall}(2003)}]{RN1539}%
  \BibitemOpen
  \bibfield  {author} {\bibinfo {author} {\bibfnamefont {B.~C.}\ \bibnamefont
  {Attwood}}\ and\ \bibinfo {author} {\bibfnamefont {C.~K.}\ \bibnamefont
  {Hall}},\ }\bibfield  {title} {\bibinfo {title} {Global phase diagram for
  monomer/dimer mixtures},\ }\href
  {https://doi.org/https://doi.org/10.1016/S0378-3812(02)00251-0} {\bibfield
  {journal} {\bibinfo  {journal} {Fluid Phase Equilibria}\ }\textbf {\bibinfo
  {volume} {204}},\ \bibinfo {pages} {85} (\bibinfo {year} {2003})}\BibitemShut
  {NoStop}%
\bibitem [{\citenamefont {Babu}\ \emph {et~al.}(2009)\citenamefont {Babu},
  \citenamefont {Gimel},\ and\ \citenamefont {Nicolai}}]{RN1271}%
  \BibitemOpen
  \bibfield  {author} {\bibinfo {author} {\bibfnamefont {S.}~\bibnamefont
  {Babu}}, \bibinfo {author} {\bibfnamefont {J.~C.}\ \bibnamefont {Gimel}},\
  and\ \bibinfo {author} {\bibfnamefont {T.}~\bibnamefont {Nicolai}},\
  }\bibfield  {title} {\bibinfo {title} {Crystallization and dynamical arrest
  of attractive hard spheres},\ }\bibfield  {journal} {\bibinfo  {journal}
  {Journal of Chemical Physics}\ }\textbf {\bibinfo {volume} {130}},\ \href
  {https://doi.org/10.1063/1.3074310} {10.1063/1.3074310} (\bibinfo {year}
  {2009})\BibitemShut {NoStop}%
\bibitem [{\citenamefont {Haxton}\ \emph {et~al.}(2015)\citenamefont {Haxton},
  \citenamefont {Hedges},\ and\ \citenamefont {Whitelam}}]{RN507}%
  \BibitemOpen
  \bibfield  {author} {\bibinfo {author} {\bibfnamefont {T.~K.}\ \bibnamefont
  {Haxton}}, \bibinfo {author} {\bibfnamefont {L.~O.}\ \bibnamefont {Hedges}},\
  and\ \bibinfo {author} {\bibfnamefont {S.}~\bibnamefont {Whitelam}},\
  }\bibfield  {title} {\bibinfo {title} {Crystallization and arrest mechanisms
  of model colloids},\ }\href {https://doi.org/10.1039/c5sm01833a} {\bibfield
  {journal} {\bibinfo  {journal} {Soft Matter}\ }\textbf {\bibinfo {volume}
  {11}},\ \bibinfo {pages} {9307} (\bibinfo {year} {2015})}\BibitemShut
  {NoStop}%
\bibitem [{\citenamefont {Prabhu}\ \emph {et~al.}(2014)\citenamefont {Prabhu},
  \citenamefont {Babu}, \citenamefont {Dolado},\ and\ \citenamefont
  {Gimel}}]{RN1273}%
  \BibitemOpen
  \bibfield  {author} {\bibinfo {author} {\bibfnamefont {A.}~\bibnamefont
  {Prabhu}}, \bibinfo {author} {\bibfnamefont {S.~B.}\ \bibnamefont {Babu}},
  \bibinfo {author} {\bibfnamefont {J.~S.}\ \bibnamefont {Dolado}},\ and\
  \bibinfo {author} {\bibfnamefont {J.~C.}\ \bibnamefont {Gimel}},\ }\bibfield
  {title} {\bibinfo {title} {Brownian cluster dynamics with short range patchy
  interactions: Its application to polymers and step-growth polymerization},\
  }\href {https://doi.org/10.1063/1.4886585} {\bibfield  {journal} {\bibinfo
  {journal} {Journal of Chemical Physics}\ }\textbf {\bibinfo {volume} {141}},\
  \bibinfo {pages} {16} (\bibinfo {year} {2014})}\BibitemShut {NoStop}%
\bibitem [{\citenamefont {Rżysko}\ \emph {et~al.}(2010)\citenamefont
  {Rżysko}, \citenamefont {Patrykiejew}, \citenamefont {Sokołowski},\ and\
  \citenamefont {Pizio}}]{RN1537}%
  \BibitemOpen
  \bibfield  {author} {\bibinfo {author} {\bibfnamefont {W.}~\bibnamefont
  {Rżysko}}, \bibinfo {author} {\bibfnamefont {A.}~\bibnamefont
  {Patrykiejew}}, \bibinfo {author} {\bibfnamefont {S.}~\bibnamefont
  {Sokołowski}},\ and\ \bibinfo {author} {\bibfnamefont {O.}~\bibnamefont
  {Pizio}},\ }\bibfield  {title} {\bibinfo {title} {Phase behavior of a
  two-dimensional and confined in slitlike pores square-shoulder, square-well
  fluid},\ }\href {https://doi.org/10.1063/1.3392744} {\bibfield  {journal}
  {\bibinfo  {journal} {The Journal of Chemical Physics}\ }\textbf {\bibinfo
  {volume} {132}},\ \bibinfo {pages} {164702} (\bibinfo {year}
  {2010})}\BibitemShut {NoStop}%
\bibitem [{\citenamefont {Serrano-Illán}\ \emph {et~al.}(2006)\citenamefont
  {Serrano-Illán}, \citenamefont {Navascués},\ and\ \citenamefont
  {Velasco}}]{RN1536}%
  \BibitemOpen
  \bibfield  {author} {\bibinfo {author} {\bibfnamefont {J.}~\bibnamefont
  {Serrano-Illán}}, \bibinfo {author} {\bibfnamefont {G.}~\bibnamefont
  {Navascués}},\ and\ \bibinfo {author} {\bibfnamefont {E.}~\bibnamefont
  {Velasco}},\ }\bibfield  {title} {\bibinfo {title} {Noncompact crystalline
  solids in the square-well potential},\ }\href
  {https://doi.org/10.1103/PhysRevE.73.011110} {\bibfield  {journal} {\bibinfo
  {journal} {Phys Rev E Stat Nonlin Soft Matter Phys}\ }\textbf {\bibinfo
  {volume} {73}},\ \bibinfo {pages} {011110} (\bibinfo {year}
  {2006})}\BibitemShut {NoStop}%
\bibitem [{\citenamefont {Sevick}\ and\ \citenamefont {Monson}(1991)}]{RN1285}%
  \BibitemOpen
  \bibfield  {author} {\bibinfo {author} {\bibfnamefont {E.~M.}\ \bibnamefont
  {Sevick}}\ and\ \bibinfo {author} {\bibfnamefont {P.~A.}\ \bibnamefont
  {Monson}},\ }\bibfield  {title} {\bibinfo {title} {Cluster integrals for
  square-well particles - application to percolation},\ }\href
  {https://doi.org/10.1063/1.459830} {\bibfield  {journal} {\bibinfo  {journal}
  {Journal of Chemical Physics}\ }\textbf {\bibinfo {volume} {94}},\ \bibinfo
  {pages} {3070} (\bibinfo {year} {1991})}\BibitemShut {NoStop}%
\bibitem [{\citenamefont {Takada}\ and\ \citenamefont
  {Hayakawa}(2018)}]{RN1283}%
  \BibitemOpen
  \bibfield  {author} {\bibinfo {author} {\bibfnamefont {S.}~\bibnamefont
  {Takada}}\ and\ \bibinfo {author} {\bibfnamefont {H.}~\bibnamefont
  {Hayakawa}},\ }\bibfield  {title} {\bibinfo {title} {Rheology of dilute
  cohesive granular gases},\ }\bibfield  {journal} {\bibinfo  {journal}
  {Physical Review E}\ }\textbf {\bibinfo {volume} {97}},\ \href
  {https://doi.org/10.1103/PhysRevE.97.042902} {10.1103/PhysRevE.97.042902}
  (\bibinfo {year} {2018})\BibitemShut {NoStop}%
\bibitem [{\citenamefont {Schnabel}\ \emph
  {et~al.}(2009{\natexlab{a}})\citenamefont {Schnabel}, \citenamefont
  {Bachmann},\ and\ \citenamefont {Janke}}]{RN1329}%
  \BibitemOpen
  \bibfield  {author} {\bibinfo {author} {\bibfnamefont {S.}~\bibnamefont
  {Schnabel}}, \bibinfo {author} {\bibfnamefont {M.}~\bibnamefont {Bachmann}},\
  and\ \bibinfo {author} {\bibfnamefont {W.}~\bibnamefont {Janke}},\ }\bibfield
   {title} {\bibinfo {title} {Elastic lennard-jones polymers meet clusters:
  Differences and similarities},\ }\bibfield  {journal} {\bibinfo  {journal}
  {Journal of Chemical Physics}\ }\textbf {\bibinfo {volume} {131}},\ \href
  {https://doi.org/10.1063/1.3223720} {10.1063/1.3223720} (\bibinfo {year}
  {2009}{\natexlab{a}})\BibitemShut {NoStop}%
\bibitem [{\citenamefont {Schnabel}\ \emph {et~al.}(2011)\citenamefont
  {Schnabel}, \citenamefont {Janke},\ and\ \citenamefont {Bachmann}}]{RN1327}%
  \BibitemOpen
  \bibfield  {author} {\bibinfo {author} {\bibfnamefont {S.}~\bibnamefont
  {Schnabel}}, \bibinfo {author} {\bibfnamefont {W.}~\bibnamefont {Janke}},\
  and\ \bibinfo {author} {\bibfnamefont {M.}~\bibnamefont {Bachmann}},\
  }\bibfield  {title} {\bibinfo {title} {Advanced multicanonical monte carlo
  methods for efficient simulations of nucleation processes of polymers},\
  }\href {https://doi.org/10.1016/j.jcp.2011.02.018} {\bibfield  {journal}
  {\bibinfo  {journal} {Journal of Computational Physics}\ }\textbf {\bibinfo
  {volume} {230}},\ \bibinfo {pages} {4454} (\bibinfo {year}
  {2011})}\BibitemShut {NoStop}%
\bibitem [{\citenamefont {Schnabel}\ \emph
  {et~al.}(2009{\natexlab{b}})\citenamefont {Schnabel}, \citenamefont {Vogel},
  \citenamefont {Bachmann},\ and\ \citenamefont {Janke}}]{RN1328}%
  \BibitemOpen
  \bibfield  {author} {\bibinfo {author} {\bibfnamefont {S.}~\bibnamefont
  {Schnabel}}, \bibinfo {author} {\bibfnamefont {T.}~\bibnamefont {Vogel}},
  \bibinfo {author} {\bibfnamefont {M.}~\bibnamefont {Bachmann}},\ and\
  \bibinfo {author} {\bibfnamefont {W.}~\bibnamefont {Janke}},\ }\bibfield
  {title} {\bibinfo {title} {Surface effects in the crystallization process of
  elastic flexible polymers},\ }\href
  {https://doi.org/10.1016/j.cplett.2009.05.052} {\bibfield  {journal}
  {\bibinfo  {journal} {Chemical Physics Letters}\ }\textbf {\bibinfo {volume}
  {476}},\ \bibinfo {pages} {201} (\bibinfo {year}
  {2009}{\natexlab{b}})}\BibitemShut {NoStop}%
\bibitem [{\citenamefont {Taylor}\ \emph {et~al.}(2009)\citenamefont {Taylor},
  \citenamefont {Paul},\ and\ \citenamefont {Binder}}]{RN1440}%
  \BibitemOpen
  \bibfield  {author} {\bibinfo {author} {\bibfnamefont {M.~P.}\ \bibnamefont
  {Taylor}}, \bibinfo {author} {\bibfnamefont {W.}~\bibnamefont {Paul}},\ and\
  \bibinfo {author} {\bibfnamefont {K.}~\bibnamefont {Binder}},\ }\bibfield
  {title} {\bibinfo {title} {Phase transitions of a single polymer chain: A
  wang–landau simulation study},\ }\href {https://doi.org/10.1063/1.3227751}
  {\bibfield  {journal} {\bibinfo  {journal} {The Journal of Chemical Physics}\
  }\textbf {\bibinfo {volume} {131}},\ \bibinfo {pages} {114907} (\bibinfo
  {year} {2009})}\BibitemShut {NoStop}%
\bibitem [{\citenamefont {Zierenberg}\ \emph {et~al.}(2016)\citenamefont
  {Zierenberg}, \citenamefont {Marenz},\ and\ \citenamefont {Janke}}]{RN1284}%
  \BibitemOpen
  \bibfield  {author} {\bibinfo {author} {\bibfnamefont {J.}~\bibnamefont
  {Zierenberg}}, \bibinfo {author} {\bibfnamefont {M.}~\bibnamefont {Marenz}},\
  and\ \bibinfo {author} {\bibfnamefont {W.}~\bibnamefont {Janke}},\ }\bibfield
   {title} {\bibinfo {title} {Dilute semiflexible polymers with attraction:
  Collapse, folding and aggregation},\ }\bibfield  {journal} {\bibinfo
  {journal} {Polymers}\ }\textbf {\bibinfo {volume} {8}},\ \href
  {https://doi.org/10.3390/polym8090333} {10.3390/polym8090333} (\bibinfo
  {year} {2016})\BibitemShut {NoStop}%
\bibitem [{\citenamefont {Woodcock}(1997)}]{RN125}%
  \BibitemOpen
  \bibfield  {author} {\bibinfo {author} {\bibfnamefont {L.~V.}\ \bibnamefont
  {Woodcock}},\ }\bibfield  {title} {\bibinfo {title} {Entropy difference
  between the face-centred cubic and hexagonal close-packed crystal
  structures},\ }\href {https://doi.org/10.1038/385141a0} {\bibfield  {journal}
  {\bibinfo  {journal} {Nature}\ }\textbf {\bibinfo {volume} {385}},\ \bibinfo
  {pages} {141} (\bibinfo {year} {1997})}\BibitemShut {NoStop}%
\bibitem [{\citenamefont {Polson}\ \emph {et~al.}(2000)\citenamefont {Polson},
  \citenamefont {Trizac}, \citenamefont {Pronk},\ and\ \citenamefont
  {Frenkel}}]{RN2011}%
  \BibitemOpen
  \bibfield  {author} {\bibinfo {author} {\bibfnamefont {J.~M.}\ \bibnamefont
  {Polson}}, \bibinfo {author} {\bibfnamefont {E.}~\bibnamefont {Trizac}},
  \bibinfo {author} {\bibfnamefont {S.}~\bibnamefont {Pronk}},\ and\ \bibinfo
  {author} {\bibfnamefont {D.}~\bibnamefont {Frenkel}},\ }\bibfield  {title}
  {\bibinfo {title} {Finite-size corrections to the free energies of
  crystalline solids},\ }\href {https://doi.org/10.1063/1.481102} {\bibfield
  {journal} {\bibinfo  {journal} {Journal of Chemical Physics}\ }\textbf
  {\bibinfo {volume} {112}},\ \bibinfo {pages} {5339} (\bibinfo {year}
  {2000})}\BibitemShut {NoStop}%
\bibitem [{\citenamefont {Bolhuis}\ \emph {et~al.}(1997)\citenamefont
  {Bolhuis}, \citenamefont {Frenkel}, \citenamefont {Mau},\ and\ \citenamefont
  {Huse}}]{RN132}%
  \BibitemOpen
  \bibfield  {author} {\bibinfo {author} {\bibfnamefont {P.~G.}\ \bibnamefont
  {Bolhuis}}, \bibinfo {author} {\bibfnamefont {D.}~\bibnamefont {Frenkel}},
  \bibinfo {author} {\bibfnamefont {S.~C.}\ \bibnamefont {Mau}},\ and\ \bibinfo
  {author} {\bibfnamefont {D.~A.}\ \bibnamefont {Huse}},\ }\bibfield  {title}
  {\bibinfo {title} {Entropy difference between crystal phases},\ }\href
  {https://doi.org/10.1038/40779} {\bibfield  {journal} {\bibinfo  {journal}
  {Nature}\ }\textbf {\bibinfo {volume} {388}},\ \bibinfo {pages} {235}
  (\bibinfo {year} {1997})}\BibitemShut {NoStop}%
\bibitem [{\citenamefont {Mau}\ and\ \citenamefont {Huse}(1999)}]{RN441}%
  \BibitemOpen
  \bibfield  {author} {\bibinfo {author} {\bibfnamefont {S.~C.}\ \bibnamefont
  {Mau}}\ and\ \bibinfo {author} {\bibfnamefont {D.~A.}\ \bibnamefont {Huse}},\
  }\bibfield  {title} {\bibinfo {title} {Stacking entropy of hard-sphere
  crystals},\ }\href {https://doi.org/10.1103/PhysRevE.59.4396} {\bibfield
  {journal} {\bibinfo  {journal} {Physical Review E}\ }\textbf {\bibinfo
  {volume} {59}},\ \bibinfo {pages} {4396} (\bibinfo {year}
  {1999})}\BibitemShut {NoStop}%
\bibitem [{\citenamefont {Noya}\ and\ \citenamefont {Almarza}(2015)}]{RN1735}%
  \BibitemOpen
  \bibfield  {author} {\bibinfo {author} {\bibfnamefont {E.~G.}\ \bibnamefont
  {Noya}}\ and\ \bibinfo {author} {\bibfnamefont {N.~G.}\ \bibnamefont
  {Almarza}},\ }\bibfield  {title} {\bibinfo {title} {Entropy of hard spheres
  in the close-packing limit},\ }\href
  {https://doi.org/10.1080/00268976.2014.982736} {\bibfield  {journal}
  {\bibinfo  {journal} {Molecular Physics}\ }\textbf {\bibinfo {volume}
  {113}},\ \bibinfo {pages} {1061} (\bibinfo {year} {2015})}\BibitemShut
  {NoStop}%
\bibitem [{\citenamefont {Ramos}\ \emph {et~al.}(2020)\citenamefont {Ramos},
  \citenamefont {Herranz}, \citenamefont {Foteinopoulou}, \citenamefont
  {Karayiannis},\ and\ \citenamefont {Laso}}]{RN1542}%
  \BibitemOpen
  \bibfield  {author} {\bibinfo {author} {\bibfnamefont {P.~M.}\ \bibnamefont
  {Ramos}}, \bibinfo {author} {\bibfnamefont {M.}~\bibnamefont {Herranz}},
  \bibinfo {author} {\bibfnamefont {K.}~\bibnamefont {Foteinopoulou}}, \bibinfo
  {author} {\bibfnamefont {N.~C.}\ \bibnamefont {Karayiannis}},\ and\ \bibinfo
  {author} {\bibfnamefont {M.}~\bibnamefont {Laso}},\ }\bibfield  {title}
  {\bibinfo {title} {Identification of local structure in 2-d and 3-d atomic
  systems through crystallographic analysis},\ }\bibfield  {journal} {\bibinfo
  {journal} {Crystals}\ }\textbf {\bibinfo {volume} {10}},\ \href
  {https://doi.org/10.3390/cryst10111008} {10.3390/cryst10111008} (\bibinfo
  {year} {2020})\BibitemShut {NoStop}%
\bibitem [{\citenamefont {Herranz}\ \emph {et~al.}(2021)\citenamefont
  {Herranz}, \citenamefont {Martínez-Fernández}, \citenamefont {Ramos},
  \citenamefont {Foteinopoulou}, \citenamefont {Karayiannis},\ and\
  \citenamefont {Laso}}]{RN1824}%
  \BibitemOpen
  \bibfield  {author} {\bibinfo {author} {\bibfnamefont {M.}~\bibnamefont
  {Herranz}}, \bibinfo {author} {\bibfnamefont {D.}~\bibnamefont
  {Martínez-Fernández}}, \bibinfo {author} {\bibfnamefont {P.~M.}\
  \bibnamefont {Ramos}}, \bibinfo {author} {\bibfnamefont {K.}~\bibnamefont
  {Foteinopoulou}}, \bibinfo {author} {\bibfnamefont {N.~C.}\ \bibnamefont
  {Karayiannis}},\ and\ \bibinfo {author} {\bibfnamefont {M.}~\bibnamefont
  {Laso}},\ }\bibfield  {title} {\bibinfo {title} {Simu-d: A
  simulator-descriptor suite for polymer-based systems under extreme
  conditions},\ }\href {https://www.mdpi.com/1422-0067/22/22/12464} {\bibfield
  {journal} {\bibinfo  {journal} {International Journal of Molecular Sciences}\
  }\textbf {\bibinfo {volume} {22}},\ \bibinfo {pages} {12464} (\bibinfo {year}
  {2021})}\BibitemShut {NoStop}%
\bibitem [{\citenamefont {Karayiannis}\ \emph {et~al.}(2002)\citenamefont
  {Karayiannis}, \citenamefont {Mavrantzas},\ and\ \citenamefont
  {Theodorou}}]{RN9}%
  \BibitemOpen
  \bibfield  {author} {\bibinfo {author} {\bibfnamefont {N.~C.}\ \bibnamefont
  {Karayiannis}}, \bibinfo {author} {\bibfnamefont {V.~G.}\ \bibnamefont
  {Mavrantzas}},\ and\ \bibinfo {author} {\bibfnamefont {D.~N.}\ \bibnamefont
  {Theodorou}},\ }\bibfield  {title} {\bibinfo {title} {A novel monte carlo
  scheme for the rapid equilibration of atomistic model polymer systems of
  precisely defined molecular architecture},\ }\bibfield  {journal} {\bibinfo
  {journal} {Physical Review Letters}\ }\textbf {\bibinfo {volume} {88}},\
  \href {https://doi.org/10.1103/PhysRevLett.88.105503}
  {10.1103/PhysRevLett.88.105503} (\bibinfo {year} {2002})\BibitemShut
  {NoStop}%
\bibitem [{\citenamefont {Karayiannis}\ and\ \citenamefont
  {Laso}(2008)}]{RN16}%
  \BibitemOpen
  \bibfield  {author} {\bibinfo {author} {\bibfnamefont {N.~C.}\ \bibnamefont
  {Karayiannis}}\ and\ \bibinfo {author} {\bibfnamefont {M.}~\bibnamefont
  {Laso}},\ }\bibfield  {title} {\bibinfo {title} {Monte carlo scheme for
  generation and relaxation of dense and nearly jammed random structures of
  freely jointed hard-sphere chains},\ }\href
  {https://doi.org/10.1021/ma702264u} {\bibfield  {journal} {\bibinfo
  {journal} {Macromolecules}\ }\textbf {\bibinfo {volume} {41}},\ \bibinfo
  {pages} {1537} (\bibinfo {year} {2008})}\BibitemShut {NoStop}%
\bibitem [{\citenamefont {Ramos}\ \emph {et~al.}(2018)\citenamefont {Ramos},
  \citenamefont {Karayiannis},\ and\ \citenamefont {Laso}}]{RN1252}%
  \BibitemOpen
  \bibfield  {author} {\bibinfo {author} {\bibfnamefont {P.~M.}\ \bibnamefont
  {Ramos}}, \bibinfo {author} {\bibfnamefont {N.~C.}\ \bibnamefont
  {Karayiannis}},\ and\ \bibinfo {author} {\bibfnamefont {M.}~\bibnamefont
  {Laso}},\ }\bibfield  {title} {\bibinfo {title} {Off-lattice simulation
  algorithms for athermal chain molecules under extreme confinement},\ }\href
  {https://doi.org/10.1016/j.jcp.2018.08.052} {\bibfield  {journal} {\bibinfo
  {journal} {Journal of Computational Physics}\ }\textbf {\bibinfo {volume}
  {375}},\ \bibinfo {pages} {918} (\bibinfo {year} {2018})}\BibitemShut
  {NoStop}%
\bibitem [{\citenamefont {Herranz}\ \emph {et~al.}(2020)\citenamefont
  {Herranz}, \citenamefont {Santiago}, \citenamefont {Foteinopoulou},
  \citenamefont {Karayiannis},\ and\ \citenamefont {Laso}}]{RN1541}%
  \BibitemOpen
  \bibfield  {author} {\bibinfo {author} {\bibfnamefont {M.}~\bibnamefont
  {Herranz}}, \bibinfo {author} {\bibfnamefont {M.}~\bibnamefont {Santiago}},
  \bibinfo {author} {\bibfnamefont {K.}~\bibnamefont {Foteinopoulou}}, \bibinfo
  {author} {\bibfnamefont {N.~C.}\ \bibnamefont {Karayiannis}},\ and\ \bibinfo
  {author} {\bibfnamefont {M.}~\bibnamefont {Laso}},\ }\bibfield  {title}
  {\bibinfo {title} {Crystal, fivefold and glass formation in clusters of
  polymers interacting with the square well potential},\ }\bibfield  {journal}
  {\bibinfo  {journal} {Polymers}\ }\textbf {\bibinfo {volume} {12}},\ \href
  {https://doi.org/10.3390/polym12051111} {10.3390/polym12051111} (\bibinfo
  {year} {2020})\BibitemShut {NoStop}%
\bibitem [{\citenamefont {Ramos}\ \emph {et~al.}(2021)\citenamefont {Ramos},
  \citenamefont {Herranz}, \citenamefont {Foteinopoulou}, \citenamefont
  {Karayiannis},\ and\ \citenamefont {Laso}}]{RN1678}%
  \BibitemOpen
  \bibfield  {author} {\bibinfo {author} {\bibfnamefont {P.~M.}\ \bibnamefont
  {Ramos}}, \bibinfo {author} {\bibfnamefont {M.}~\bibnamefont {Herranz}},
  \bibinfo {author} {\bibfnamefont {K.}~\bibnamefont {Foteinopoulou}}, \bibinfo
  {author} {\bibfnamefont {N.~C.}\ \bibnamefont {Karayiannis}},\ and\ \bibinfo
  {author} {\bibfnamefont {M.}~\bibnamefont {Laso}},\ }\bibfield  {title}
  {\bibinfo {title} {Entropy-driven heterogeneous crystallization of
  hard-sphere chains under unidimensional confinement},\ }\bibfield  {journal}
  {\bibinfo  {journal} {Polymers}\ }\textbf {\bibinfo {volume} {13}},\ \href
  {https://doi.org/10.3390/polym13091352} {10.3390/polym13091352} (\bibinfo
  {year} {2021})\BibitemShut {NoStop}%
\bibitem [{\citenamefont {Herranz}\ \emph {et~al.}(2022)\citenamefont
  {Herranz}, \citenamefont {Foteinopoulou}, \citenamefont {Karayiannis},\ and\
  \citenamefont {Laso}}]{RN1894}%
  \BibitemOpen
  \bibfield  {author} {\bibinfo {author} {\bibfnamefont {M.}~\bibnamefont
  {Herranz}}, \bibinfo {author} {\bibfnamefont {K.}~\bibnamefont
  {Foteinopoulou}}, \bibinfo {author} {\bibfnamefont {N.~C.}\ \bibnamefont
  {Karayiannis}},\ and\ \bibinfo {author} {\bibfnamefont {M.}~\bibnamefont
  {Laso}},\ }\bibfield  {title} {\bibinfo {title} {Polymorphism and perfection
  in crystallization of hard sphere polymers},\ }\bibfield  {journal} {\bibinfo
   {journal} {Polymers}\ }\textbf {\bibinfo {volume} {14}},\ \href
  {https://doi.org/10.3390/polym14204435} {10.3390/polym14204435} (\bibinfo
  {year} {2022})\BibitemShut {NoStop}%
\bibitem [{\citenamefont {Martinez-Fernandez}\ \emph
  {et~al.}(2023)\citenamefont {Martinez-Fernandez}, \citenamefont {Herranz},
  \citenamefont {Foteinopoulou}, \citenamefont {Karayiannis},\ and\
  \citenamefont {Laso}}]{RN2010}%
  \BibitemOpen
  \bibfield  {author} {\bibinfo {author} {\bibfnamefont {D.}~\bibnamefont
  {Martinez-Fernandez}}, \bibinfo {author} {\bibfnamefont {M.}~\bibnamefont
  {Herranz}}, \bibinfo {author} {\bibfnamefont {K.}~\bibnamefont
  {Foteinopoulou}}, \bibinfo {author} {\bibfnamefont {N.~C.}\ \bibnamefont
  {Karayiannis}},\ and\ \bibinfo {author} {\bibfnamefont {M.}~\bibnamefont
  {Laso}},\ }\bibfield  {title} {\bibinfo {title} {Local and global order in
  dense packings of semi-flexible polymers of hard spheres},\ }\bibfield
  {journal} {\bibinfo  {journal} {Polymers}\ }\textbf {\bibinfo {volume}
  {15}},\ \href {https://doi.org/10.3390/polym15030551} {10.3390/polym15030551}
  (\bibinfo {year} {2023})\BibitemShut {NoStop}%
\bibitem [{\citenamefont {Ester}\ \emph {et~al.}(1996)\citenamefont {Ester},
  \citenamefont {Kriegel}, \citenamefont {Sander},\ and\ \citenamefont
  {Xu}}]{confkddEsterKSX96}%
  \BibitemOpen
  \bibfield  {author} {\bibinfo {author} {\bibfnamefont {M.}~\bibnamefont
  {Ester}}, \bibinfo {author} {\bibfnamefont {H.-P.}\ \bibnamefont {Kriegel}},
  \bibinfo {author} {\bibfnamefont {J.}~\bibnamefont {Sander}},\ and\ \bibinfo
  {author} {\bibfnamefont {X.}~\bibnamefont {Xu}},\ }\bibfield  {title}
  {\bibinfo {title} {A density-based algorithm for discovering clusters in
  large spatial databases with noise.},\ }in\ \href
  {httpdblp.uni-trier.dedbconfkddkdd96.html#EsterKSX96} {\emph {\bibinfo
  {booktitle} {KDD}}},\ \bibinfo {editor} {edited by\ \bibinfo {editor}
  {\bibfnamefont {E.}~\bibnamefont {Simoudis}}, \bibinfo {editor}
  {\bibfnamefont {J.}~\bibnamefont {Han}},\ and\ \bibinfo {editor}
  {\bibfnamefont {U.~M.}\ \bibnamefont {Fayyad}}}\ (\bibinfo  {publisher} {AAAI
  Press},\ \bibinfo {year} {1996})\ pp.\ \bibinfo {pages}
  {226--231}\BibitemShut {NoStop}%
\bibitem [{\citenamefont {Schubert}\ \emph {et~al.}(2017)\citenamefont
  {Schubert}, \citenamefont {Sander}, \citenamefont {Ester}, \citenamefont
  {Kriegel},\ and\ \citenamefont {Xu}}]{RN2012}%
  \BibitemOpen
  \bibfield  {author} {\bibinfo {author} {\bibfnamefont {E.}~\bibnamefont
  {Schubert}}, \bibinfo {author} {\bibfnamefont {J.}~\bibnamefont {Sander}},
  \bibinfo {author} {\bibfnamefont {M.}~\bibnamefont {Ester}}, \bibinfo
  {author} {\bibfnamefont {H.~P.}\ \bibnamefont {Kriegel}},\ and\ \bibinfo
  {author} {\bibfnamefont {X.~W.}\ \bibnamefont {Xu}},\ }\bibfield  {title}
  {\bibinfo {title} {Dbscan revisited, revisited: Why and how you should
  (still) use dbscan},\ }\bibfield  {journal} {\bibinfo  {journal} {Acm
  Transactions on Database Systems}\ }\textbf {\bibinfo {volume} {42}},\ \href
  {https://doi.org/10.1145/3068335} {10.1145/3068335} (\bibinfo {year}
  {2017})\BibitemShut {NoStop}%
\bibitem [{\citenamefont {Pedrosa}\ \emph {et~al.}(2023)\citenamefont
  {Pedrosa}, \citenamefont {Martinez-Fernandez}, \citenamefont {Herranz},
  \citenamefont {Foteinopoulou}, \citenamefont {Karayiannis},\ and\
  \citenamefont {Laso}}]{ClaraP}%
  \BibitemOpen
  \bibfield  {author} {\bibinfo {author} {\bibfnamefont {P.}~\bibnamefont
  {Pedrosa}}, \bibinfo {author} {\bibfnamefont {D.}~\bibnamefont
  {Martinez-Fernandez}}, \bibinfo {author} {\bibfnamefont {M.}~\bibnamefont
  {Herranz}}, \bibinfo {author} {\bibfnamefont {K.}~\bibnamefont
  {Foteinopoulou}}, \bibinfo {author} {\bibfnamefont {N.~C.}\ \bibnamefont
  {Karayiannis}},\ and\ \bibinfo {author} {\bibfnamefont {M.}~\bibnamefont
  {Laso}},\ }\bibfield  {title} {\bibinfo {title} {Densest packing of flexible
  polymers in 2-d films},\ }\bibfield  {journal} {\bibinfo  {journal} {Journal
  of Chemical Physics}\ }\textbf {\bibinfo {volume} {158}},\ \href
  {https://doi.org/10.1063/5.0137115} {10.1063/5.0137115} (\bibinfo {year}
  {2023})\BibitemShut {NoStop}%
\bibitem [{\citenamefont {Karayiannis}\ \emph {et~al.}(2009)\citenamefont
  {Karayiannis}, \citenamefont {Foteinopoulou},\ and\ \citenamefont
  {Laso}}]{RN10}%
  \BibitemOpen
  \bibfield  {author} {\bibinfo {author} {\bibfnamefont {N.~C.}\ \bibnamefont
  {Karayiannis}}, \bibinfo {author} {\bibfnamefont {K.}~\bibnamefont
  {Foteinopoulou}},\ and\ \bibinfo {author} {\bibfnamefont {M.}~\bibnamefont
  {Laso}},\ }\bibfield  {title} {\bibinfo {title} {The characteristic
  crystallographic element norm: A descriptor of local structure in atomistic
  and particulate systems},\ }\bibfield  {journal} {\bibinfo  {journal}
  {Journal of Chemical Physics}\ }\textbf {\bibinfo {volume} {130}},\ \href
  {https://doi.org/10.1063/1.3077294} {10.1063/1.3077294} (\bibinfo {year}
  {2009})\BibitemShut {NoStop}%
\bibitem [{\citenamefont {Humphrey}\ \emph {et~al.}(1996)\citenamefont
  {Humphrey}, \citenamefont {Dalke},\ and\ \citenamefont {Schulten}}]{RN250}%
  \BibitemOpen
  \bibfield  {author} {\bibinfo {author} {\bibfnamefont {W.}~\bibnamefont
  {Humphrey}}, \bibinfo {author} {\bibfnamefont {A.}~\bibnamefont {Dalke}},\
  and\ \bibinfo {author} {\bibfnamefont {K.}~\bibnamefont {Schulten}},\
  }\bibfield  {title} {\bibinfo {title} {Vmd: Visual molecular dynamics},\
  }\href {https://doi.org/10.1016/0263-7855(96)00018-5} {\bibfield  {journal}
  {\bibinfo  {journal} {Journal of Molecular Graphics \& Modelling}\ }\textbf
  {\bibinfo {volume} {14}},\ \bibinfo {pages} {33} (\bibinfo {year}
  {1996})}\BibitemShut {NoStop}%
\bibitem [{\citenamefont {Herranz}\ \emph {et~al.}(2023)\citenamefont
  {Herranz}, \citenamefont {Benito}, \citenamefont {Foteinopoulou},
  \citenamefont {Karayiannis},\ and\ \citenamefont {Laso}}]{MiguelH}%
  \BibitemOpen
  \bibfield  {author} {\bibinfo {author} {\bibfnamefont {M.}~\bibnamefont
  {Herranz}}, \bibinfo {author} {\bibfnamefont {J.}~\bibnamefont {Benito}},
  \bibinfo {author} {\bibfnamefont {K.}~\bibnamefont {Foteinopoulou}}, \bibinfo
  {author} {\bibfnamefont {N.~C.}\ \bibnamefont {Karayiannis}},\ and\ \bibinfo
  {author} {\bibfnamefont {M.}~\bibnamefont {Laso}},\ }\bibfield  {title}
  {\bibinfo {title} {Polymorph stability and free energy of crystallization of
  freely-jointed polymers of hard spheres},\ }\bibfield  {journal} {\bibinfo
  {journal} {Polymers}\ }\textbf {\bibinfo {volume} {15}},\ \href
  {https://doi.org/10.3390/polym15061335} {10.3390/polym15061335} (\bibinfo
  {year} {2023})\BibitemShut {NoStop}%
\bibitem [{Sup()}]{Suppl}%
  \BibitemOpen
  \href@noop {} {\bibinfo {title} {See supplemental material at
  https://journals.aps.org/}}\BibitemShut {NoStop}%
\bibitem [{\citenamefont {Frank}\ and\ \citenamefont {Kasper}(1958)}]{RN1951}%
  \BibitemOpen
  \bibfield  {author} {\bibinfo {author} {\bibfnamefont {F.~C.}\ \bibnamefont
  {Frank}}\ and\ \bibinfo {author} {\bibfnamefont {J.~S.}\ \bibnamefont
  {Kasper}},\ }\bibfield  {title} {\bibinfo {title} {Complex alloy structures
  regarded as sphere packings .1. definitions and basic principles},\ }\href
  {https://doi.org/10.1107/s0365110x58000487} {\bibfield  {journal} {\bibinfo
  {journal} {Acta Crystallographica}\ }\textbf {\bibinfo {volume} {11}},\
  \bibinfo {pages} {184} (\bibinfo {year} {1958})}\BibitemShut {NoStop}%
\bibitem [{\citenamefont {Frank}\ and\ \citenamefont {Kasper}(1959)}]{RN1952}%
  \BibitemOpen
  \bibfield  {author} {\bibinfo {author} {\bibfnamefont {F.~C.}\ \bibnamefont
  {Frank}}\ and\ \bibinfo {author} {\bibfnamefont {J.~S.}\ \bibnamefont
  {Kasper}},\ }\bibfield  {title} {\bibinfo {title} {Complex alloy structures
  regarded as sphere packing .2. analysis and classification of representative
  structures},\ }\href {https://doi.org/10.1107/s0365110x59001499} {\bibfield
  {journal} {\bibinfo  {journal} {Acta Crystallographica}\ }\textbf {\bibinfo
  {volume} {12}},\ \bibinfo {pages} {483} (\bibinfo {year} {1959})}\BibitemShut
  {NoStop}%
\bibitem [{\citenamefont {Huang}\ \emph {et~al.}(2018)\citenamefont {Huang},
  \citenamefont {Yue}, \citenamefont {Wang}, \citenamefont {Hsu}, \citenamefont
  {Wang},\ and\ \citenamefont {Cheng}}]{RN1980}%
  \BibitemOpen
  \bibfield  {author} {\bibinfo {author} {\bibfnamefont {M.~J.}\ \bibnamefont
  {Huang}}, \bibinfo {author} {\bibfnamefont {K.}~\bibnamefont {Yue}}, \bibinfo
  {author} {\bibfnamefont {J.}~\bibnamefont {Wang}}, \bibinfo {author}
  {\bibfnamefont {C.~H.}\ \bibnamefont {Hsu}}, \bibinfo {author} {\bibfnamefont
  {L.~G.}\ \bibnamefont {Wang}},\ and\ \bibinfo {author} {\bibfnamefont
  {S.~Z.~D.}\ \bibnamefont {Cheng}},\ }\bibfield  {title} {\bibinfo {title}
  {Frank-kasper and related quasicrystal spherical phases in macromolecules},\
  }\href {https://doi.org/10.1007/s11426-017-9140-8} {\bibfield  {journal}
  {\bibinfo  {journal} {Science China-Chemistry}\ }\textbf {\bibinfo {volume}
  {61}},\ \bibinfo {pages} {33} (\bibinfo {year} {2018})}\BibitemShut {NoStop}%
\bibitem [{\citenamefont {Ungar}\ and\ \citenamefont {Zeng}(2005)}]{RN1981}%
  \BibitemOpen
  \bibfield  {author} {\bibinfo {author} {\bibfnamefont {G.}~\bibnamefont
  {Ungar}}\ and\ \bibinfo {author} {\bibfnamefont {X.~B.}\ \bibnamefont
  {Zeng}},\ }\bibfield  {title} {\bibinfo {title} {Frank-kasper,
  quasicrystalline and related phases in liquid crystals},\ }\href
  {https://doi.org/10.1039/b502443a} {\bibfield  {journal} {\bibinfo  {journal}
  {Soft Matter}\ }\textbf {\bibinfo {volume} {1}},\ \bibinfo {pages} {95}
  (\bibinfo {year} {2005})}\BibitemShut {NoStop}%
\bibitem [{\citenamefont {Huang}\ \emph {et~al.}(2015)\citenamefont {Huang},
  \citenamefont {Hsu}, \citenamefont {Wang}, \citenamefont {Mei}, \citenamefont
  {Dong}, \citenamefont {Li}, \citenamefont {Li}, \citenamefont {Liu},
  \citenamefont {Zhang}, \citenamefont {Aida}, \citenamefont {Zhang},
  \citenamefont {Yue},\ and\ \citenamefont {Cheng}}]{RN1982}%
  \BibitemOpen
  \bibfield  {author} {\bibinfo {author} {\bibfnamefont {M.~J.}\ \bibnamefont
  {Huang}}, \bibinfo {author} {\bibfnamefont {C.~H.}\ \bibnamefont {Hsu}},
  \bibinfo {author} {\bibfnamefont {J.}~\bibnamefont {Wang}}, \bibinfo {author}
  {\bibfnamefont {S.}~\bibnamefont {Mei}}, \bibinfo {author} {\bibfnamefont
  {X.~H.}\ \bibnamefont {Dong}}, \bibinfo {author} {\bibfnamefont {Y.~W.}\
  \bibnamefont {Li}}, \bibinfo {author} {\bibfnamefont {M.~X.}\ \bibnamefont
  {Li}}, \bibinfo {author} {\bibfnamefont {H.}~\bibnamefont {Liu}}, \bibinfo
  {author} {\bibfnamefont {W.}~\bibnamefont {Zhang}}, \bibinfo {author}
  {\bibfnamefont {T.~Z.}\ \bibnamefont {Aida}}, \bibinfo {author}
  {\bibfnamefont {W.~B.}\ \bibnamefont {Zhang}}, \bibinfo {author}
  {\bibfnamefont {K.}~\bibnamefont {Yue}},\ and\ \bibinfo {author}
  {\bibfnamefont {S.~Z.~D.}\ \bibnamefont {Cheng}},\ }\bibfield  {title}
  {\bibinfo {title} {Selective assemblies of giant tetrahedra via precisely
  controlled positional interactions},\ }\href
  {https://doi.org/10.1126/science.aaa2421} {\bibfield  {journal} {\bibinfo
  {journal} {Science}\ }\textbf {\bibinfo {volume} {348}},\ \bibinfo {pages}
  {424} (\bibinfo {year} {2015})}\BibitemShut {NoStop}%
\bibitem [{\citenamefont {Chang}\ and\ \citenamefont {Bates}(2020)}]{RN1983}%
  \BibitemOpen
  \bibfield  {author} {\bibinfo {author} {\bibfnamefont {A.~B.}\ \bibnamefont
  {Chang}}\ and\ \bibinfo {author} {\bibfnamefont {F.~S.}\ \bibnamefont
  {Bates}},\ }\bibfield  {title} {\bibinfo {title} {Impact of architectural
  asymmetry on frank-kasper phase formation in block polymer melts},\ }\href
  {https://doi.org/10.1021/acsnano.0c03846} {\bibfield  {journal} {\bibinfo
  {journal} {Acs Nano}\ }\textbf {\bibinfo {volume} {14}},\ \bibinfo {pages}
  {11463} (\bibinfo {year} {2020})}\BibitemShut {NoStop}%
\bibitem [{\citenamefont {Su}\ \emph {et~al.}(2020)\citenamefont {Su},
  \citenamefont {Zhang}, \citenamefont {Yan}, \citenamefont {Guo},
  \citenamefont {Huang}, \citenamefont {Shan}, \citenamefont {Liu},
  \citenamefont {Liu}, \citenamefont {Huang},\ and\ \citenamefont
  {Cheng}}]{RN1984}%
  \BibitemOpen
  \bibfield  {author} {\bibinfo {author} {\bibfnamefont {Z.~B.}\ \bibnamefont
  {Su}}, \bibinfo {author} {\bibfnamefont {R.~M.}\ \bibnamefont {Zhang}},
  \bibinfo {author} {\bibfnamefont {X.~Y.}\ \bibnamefont {Yan}}, \bibinfo
  {author} {\bibfnamefont {Q.~Y.}\ \bibnamefont {Guo}}, \bibinfo {author}
  {\bibfnamefont {J.~H.}\ \bibnamefont {Huang}}, \bibinfo {author}
  {\bibfnamefont {W.~P.}\ \bibnamefont {Shan}}, \bibinfo {author}
  {\bibfnamefont {Y.~C.}\ \bibnamefont {Liu}}, \bibinfo {author} {\bibfnamefont
  {T.}~\bibnamefont {Liu}}, \bibinfo {author} {\bibfnamefont {M.~J.}\
  \bibnamefont {Huang}},\ and\ \bibinfo {author} {\bibfnamefont {S.~Z.~D.}\
  \bibnamefont {Cheng}},\ }\bibfield  {title} {\bibinfo {title} {The role of
  architectural engineering in macromolecular self-assemblies via non-covalent
  interactions: A molecular lego approach},\ }\bibfield  {journal} {\bibinfo
  {journal} {Progress in Polymer Science}\ }\textbf {\bibinfo {volume} {103}},\
  \href {https://doi.org/10.1016/j.progpolymsci.2020.101230}
  {10.1016/j.progpolymsci.2020.101230} (\bibinfo {year} {2020})\BibitemShut
  {NoStop}%
\bibitem [{\citenamefont {Han}\ and\ \citenamefont {Che}(2013)}]{RN1985}%
  \BibitemOpen
  \bibfield  {author} {\bibinfo {author} {\bibfnamefont {L.}~\bibnamefont
  {Han}}\ and\ \bibinfo {author} {\bibfnamefont {S.~N.}\ \bibnamefont {Che}},\
  }\bibfield  {title} {\bibinfo {title} {Anionic surfactant templated
  mesoporous silicas (amss)},\ }\href {https://doi.org/10.1039/c2cs35297d}
  {\bibfield  {journal} {\bibinfo  {journal} {Chemical Society Reviews}\
  }\textbf {\bibinfo {volume} {42}},\ \bibinfo {pages} {3740} (\bibinfo {year}
  {2013})}\BibitemShut {NoStop}%
\bibitem [{\citenamefont {Hudson}\ \emph {et~al.}(1997)\citenamefont {Hudson},
  \citenamefont {Jung}, \citenamefont {Percec}, \citenamefont {Cho},
  \citenamefont {Johansson}, \citenamefont {Ungar},\ and\ \citenamefont
  {Balagurusamy}}]{RN1986}%
  \BibitemOpen
  \bibfield  {author} {\bibinfo {author} {\bibfnamefont {S.~D.}\ \bibnamefont
  {Hudson}}, \bibinfo {author} {\bibfnamefont {H.~T.}\ \bibnamefont {Jung}},
  \bibinfo {author} {\bibfnamefont {V.}~\bibnamefont {Percec}}, \bibinfo
  {author} {\bibfnamefont {W.~D.}\ \bibnamefont {Cho}}, \bibinfo {author}
  {\bibfnamefont {G.}~\bibnamefont {Johansson}}, \bibinfo {author}
  {\bibfnamefont {G.}~\bibnamefont {Ungar}},\ and\ \bibinfo {author}
  {\bibfnamefont {V.~S.~K.}\ \bibnamefont {Balagurusamy}},\ }\bibfield  {title}
  {\bibinfo {title} {Direct visualization of individual cylindrical and
  spherical supramolecular dendrimers},\ }\href
  {https://doi.org/10.1126/science.278.5337.449} {\bibfield  {journal}
  {\bibinfo  {journal} {Science}\ }\textbf {\bibinfo {volume} {278}},\ \bibinfo
  {pages} {449} (\bibinfo {year} {1997})}\BibitemShut {NoStop}%
\bibitem [{\citenamefont {Shevchenko}\ \emph {et~al.}(2006)\citenamefont
  {Shevchenko}, \citenamefont {Talapin}, \citenamefont {Kotov}, \citenamefont
  {O'Brien},\ and\ \citenamefont {Murray}}]{RN1987}%
  \BibitemOpen
  \bibfield  {author} {\bibinfo {author} {\bibfnamefont {E.~V.}\ \bibnamefont
  {Shevchenko}}, \bibinfo {author} {\bibfnamefont {D.~V.}\ \bibnamefont
  {Talapin}}, \bibinfo {author} {\bibfnamefont {N.~A.}\ \bibnamefont {Kotov}},
  \bibinfo {author} {\bibfnamefont {S.}~\bibnamefont {O'Brien}},\ and\ \bibinfo
  {author} {\bibfnamefont {C.~B.}\ \bibnamefont {Murray}},\ }\bibfield  {title}
  {\bibinfo {title} {Structural diversity in binary nanoparticle
  superlattices},\ }\href {https://doi.org/10.1038/nature04414} {\bibfield
  {journal} {\bibinfo  {journal} {Nature}\ }\textbf {\bibinfo {volume} {439}},\
  \bibinfo {pages} {55} (\bibinfo {year} {2006})}\BibitemShut {NoStop}%
\bibitem [{\citenamefont {Takagi}\ and\ \citenamefont
  {Yamamoto}(2019)}]{RN1988}%
  \BibitemOpen
  \bibfield  {author} {\bibinfo {author} {\bibfnamefont {H.}~\bibnamefont
  {Takagi}}\ and\ \bibinfo {author} {\bibfnamefont {K.}~\bibnamefont
  {Yamamoto}},\ }\bibfield  {title} {\bibinfo {title} {Phase boundary of
  frank-kasper sigma phase in phase diagrams of binary mixtures of block
  copolymers and homopolymers},\ }\href
  {https://doi.org/10.1021/acs.macromol.8b02356} {\bibfield  {journal}
  {\bibinfo  {journal} {Macromolecules}\ }\textbf {\bibinfo {volume} {52}},\
  \bibinfo {pages} {2007} (\bibinfo {year} {2019})}\BibitemShut {NoStop}%
\bibitem [{\citenamefont {Chen}\ \emph {et~al.}(2022)\citenamefont {Chen},
  \citenamefont {Huang}, \citenamefont {Chen},\ and\ \citenamefont
  {Chen}}]{RN1989}%
  \BibitemOpen
  \bibfield  {author} {\bibinfo {author} {\bibfnamefont {M.~Z.}\ \bibnamefont
  {Chen}}, \bibinfo {author} {\bibfnamefont {Y.~T.}\ \bibnamefont {Huang}},
  \bibinfo {author} {\bibfnamefont {C.~Y.}\ \bibnamefont {Chen}},\ and\
  \bibinfo {author} {\bibfnamefont {H.~L.}\ \bibnamefont {Chen}},\ }\bibfield
  {title} {\bibinfo {title} {Accessing the frank-kasper sigma phase of block
  copolymer with small conformational asymmetry via selective solvent
  solubilization in the micellar corona},\ }\href
  {https://doi.org/10.1021/acs.macromol.2c01757} {\bibfield  {journal}
  {\bibinfo  {journal} {Macromolecules}\ }\textbf {\bibinfo {volume} {55}},\
  \bibinfo {pages} {10812} (\bibinfo {year} {2022})}\BibitemShut {NoStop}%
\bibitem [{\citenamefont {Lindquist}\ \emph {et~al.}(2018)\citenamefont
  {Lindquist}, \citenamefont {Jadrich}, \citenamefont {Pineros},\ and\
  \citenamefont {Truskett}}]{RN1990}%
  \BibitemOpen
  \bibfield  {author} {\bibinfo {author} {\bibfnamefont {B.~A.}\ \bibnamefont
  {Lindquist}}, \bibinfo {author} {\bibfnamefont {R.~B.}\ \bibnamefont
  {Jadrich}}, \bibinfo {author} {\bibfnamefont {W.~D.}\ \bibnamefont
  {Pineros}},\ and\ \bibinfo {author} {\bibfnamefont {T.~M.}\ \bibnamefont
  {Truskett}},\ }\bibfield  {title} {\bibinfo {title} {Inverse design of
  self-assembling frank-kasper phases and insights into emergent
  quasicrystals},\ }\href {https://doi.org/10.1021/acs.jpcb.7b11841} {\bibfield
   {journal} {\bibinfo  {journal} {Journal of Physical Chemistry B}\ }\textbf
  {\bibinfo {volume} {122}},\ \bibinfo {pages} {5547} (\bibinfo {year}
  {2018})}\BibitemShut {NoStop}%
\bibitem [{\citenamefont {Rycroft}(2009)}]{RN1543}%
  \BibitemOpen
  \bibfield  {author} {\bibinfo {author} {\bibfnamefont {C.~H.}\ \bibnamefont
  {Rycroft}},\ }\bibfield  {title} {\bibinfo {title} {Voro++: A
  three-dimensional voronoi cell library in c++},\ }\href
  {https://doi.org/10.1063/1.3215722} {\bibfield  {journal} {\bibinfo
  {journal} {Chaos: An Interdisciplinary Journal of Nonlinear Science}\
  }\textbf {\bibinfo {volume} {19}},\ \bibinfo {pages} {041111} (\bibinfo
  {year} {2009})}\BibitemShut {NoStop}%
\end{thebibliography}%
%please don't delete the next line; toggle the line above
%\bibliography{../Main_bibbkp.bib}
%
\end{document}